\documentclass[]{interact}

\usepackage{epstopdf}% To incorporate .eps illustrations using PDFLaTeX, etc.
\usepackage[caption=false]{subfig}% Support for small, `sub' figures and tables
\usepackage{float}
\usepackage{amssymb,amsmath}
\usepackage{amsmath}
\usepackage[title]{appendix}

\DeclareMathOperator{\Prob}{\mbox{P}}
\usepackage{bm}
\usepackage{rotating}
\usepackage{multirow}
\usepackage{hyperref}
\usepackage{float}
\usepackage{natbib}
\usepackage{algorithm}
\usepackage{algpseudocode}
\usepackage{diagbox}
\usepackage{threeparttable}
\numberwithin{equation}{section}
\newcommand{\B}{\boldsymbol}

\begin{document}
\let\today\relax
%\articletype{Preprint}% Specify the article type or omit as appropriate

\title{Modelling extreme claims via composite models and threshold selection methods}

\author{
\name{Yinzhi Wang\textsuperscript{a}\thanks{CONTACT Yinzhi Wang. Email: yinzhiw@math.uio.no}, Ingrid Hob\ae k Haff\textsuperscript{a} and Arne Huseby\textsuperscript{a}}
\affil{\textsuperscript{a}Department of Mathematics, University of Oslo, Postboks 1053 Blindern, 0316 Oslo, Norway}
}

\maketitle
\begin{abstract}
The existence of large and extreme claims of a non-life insurance portfolio influences the ability of (re)insurers to estimate the reserve. The excess over-threshold method provides a way to capture and model the typical behaviour of insurance claim data. This paper discusses several composite models with commonly used bulk distributions, combined with a 2-parameter Pareto distribution above the threshold. We have explored how several threshold selection methods perform when estimating the reserve as well as the effect of the choice of bulk distribution, with varying sample size and tail properties. To investigate this, a simulation study has been performed. Our study shows that when data are sufficient, the square root rule has the overall best performance in term of the quality of the reserve estimate. The second best is the exponentiality test, especially when the right tail of the data is extreme. As the sample size becomes small, the simultaneous estimation has the best performance. Further, the influence of the choice of bulk distribution seems to be rather large, especially when the distribution is heavy-tailed. Moreover, it shows that the empirical estimate of $p_{\leq b}$, the probability that a claim is below the threshold, is more robust than the theoretical one.  
\end{abstract}

\begin{keywords}
Loss distributions, Excess over-threshold method, Threshold selection, Risk measures, Monte Carlo simulation
\end{keywords}

\section{Introduction}\label{sec:introduction}

In non-life insurance, a few large losses from a portfolio of policies may represent a major proportion of the total cost. Such extreme events are of great interest for actuaries as they play an important role in decision making procedure for insurers, for instance for calculating premia, measuring tail risk and finding optimal reinsurance schemes. The classic stochastic model for the total amount the insurer pays due to claims over a defined period of time is the collective risk model, where the claim number $\mathcal{N}$ and the individual losses $Z_i$ are modelled separately by the claim frequency and the claim severity distributions, respectively. The total claim loss  $\mathcal{X}$ is then given by
$$\mathcal{X}= \sum_{i=1}^{\mathcal{N}}Z_i.
$$

It is common to assume that the claim severities are identically distributed and independent of each other and of the claim number \citep{kaas2001, klugman2012loss}. In particular, the assumption that claim severities are identically distributed is unrealistic. However, the heterogeneity between the claim severities for different policies will be averaged out when we consider the aggregate loss, so that this assumption is adequate in this setting.  

An important application of the collective risk model is the estimation of the claim reserve, or solvency capital \citep{bolviken_2014}, which is a quantile far out in the tail of the distribution of $\mathcal{X}$, given by
$$ \Pr(\mathcal{X}\geq q_\epsilon)=\epsilon,$$
where $\epsilon$ typically is 0.01 or 0.005. The reserve tends to be much more influenced by the claim severity than by the claim frequency distribution \citep{klugman2012loss}. The focus in this paper is therefore on the claim severity model. For the claim frequency, we simply assume the classic Poisson distribution with fixed intensity for all policies. Although it is a simplification, it is sufficient in this setting. 

Choosing a suitable claim severity distribution is vital, especially in the presence of extreme values. Claim severities in non-life insurance are positive random variables, that tend to follow a skew distribution with a heavier right tail. Several heavy tailed models have been used in non-life insurance, including the Gamma, the log-normal, the log-Gamma, the Weibull, the Pareto \citep{kleiber, klugman2012loss} and the extended Pareto \citep{bolviken_2014}, to name a few. These models may fit small and moderate claims well, but when the main interest is in the tail of total loss distribution, such a model may easily underestimate large claims severely. Therefore it is essential to have a good model for both moderate and large claims. Extreme value theory (EVT) and the properties of the Generalized Pareto distribution(GPD) allow us to construct a composite model which consists of two parts; a sub-threshold, or bulk distribution and an over-threshold distribution, modelling claims below and above the threshold $b$, respectively \citep{bolviken_2014}. This method for constructing composite models is known as the excess over threshold (EOT) or peaks over threshold (POT) approach \citep{smithRL} and is commonly used within non-life insurance.

One of the main concerns of EOT is the selection of an appropriate threshold for a given data set. Many approaches for threshold estimation in extreme value applications have been developed. The traditional fixed threshold estimation method uses graphical diagnostics to make a choice of threshold visually. The mean residual life (or mean excess) plot \citep{Davison.Smith1990}, which based on the mean excess function, is a quite common tool in the field of insurance. \cite{GPD} use a sequential Mann-Kendall test, known as the Gertensgarbe Plot proposed by \cite{gertensgarbe1989}. Other commonly used graphical diagnostics can be found in  \cite{Coles2001}, including threshold stability, quantile and return level plots. 

There have also been proposed a number of heuristic methods for threshold selection, which include the upper $10\%$  rule of \cite{DuMouchel}, the square root rule \citep{Ferreira2003} and the empirical rule \citep{LORETAN1994211}, all outlined in \cite{Scarrott_areview}. These approaches have no theoretical justification, but are easy to implement and frequently used in practice. 

Another alternative is automatic threshold selection with the aim of balancing bias and variance. The most common methods of this type are based on minimization of some kind of mean squared error (MSE) estimate, and are discussed in \cite{EVA2015}. \cite{HALL1990177} and \cite{Gomes2001} use the bootstrap in the threshold selection, whereas \cite{Guillou2001} and \cite{DREES1998149} propose a bias reduction procedure for choosing the optimal threshold. 

In the above mentioned methods, the threshold is selected in a first step. Then the parameters of the distributions above and below the threshold are estimated keeping the threshold fixed. \cite{actuarial} propose a composite model with a corresponding estimation method, that allows for a simultaneous estimation of the threshold and other model parameters. Their model is generalised by \cite{scollnik2007}. Yet another approach is to use a threshold mixing model, in order to take the threshold uncertainty into account, as in \cite{Tancredi2006}. Further, \cite{pigeon2011composite} develop the composite model of \cite{actuarial} to allow for heterogeneity of the threshold and let it vary among data.

The question is what the optimal threshold selection method for non-life insurance claim data is. In particular, since limitation of data is common within certain business lines and for young companies, inceasing parameter uncertainty with decreasing amount of available data should be considered. The crucial point here is not the error in the estimated parameters themselves, but rather how different threshold selection methods affect the measure of primary interest, namely the claim reserve and how they perform as the sample size decreases. Most of the literature on threshold selection
focuses on the fit of the distribution above the threshold. As the reserve also depends on the distribution below the threshold, the so-called bulk distribution, we also want to investigate how much the estimation of the reserve is affected by the choice of bulk distribution. In order to investigate this, we will perform a large simulation study. 

The rest of the paper is organized as follows. Section \ref{sec:compositemodels} gives an overview of the considered composite models. Then, the threshold selection methods compared in the simulation study are presented in Section \ref{sec:thresholdselect}. Further, the simulation study and its results are given in Section \ref{sec:simstudy}. In Section \ref{sec:realdata}, we apply the methods from Section \ref{sec:thresholdselect} to a real data set. Finally, Section \ref{sec:conclusion} provides a discussion and some concluding remarks.

\section{Composite models}\label{sec:compositemodels}

The usual way of modelling claim severity data that contain extreme values is to use a compsite distribution, where claim sizes below a certain threshold $b$ are assumed to be ``normal", and hence follow one of the common claim size distributions, right truncated at $b$, and claim sizes above the threshold are assumed to be extreme. Let $f_1$ and $f_2$ be two probability density functions(pdf's) concentrated on $[0, \infty)$, and let $b > 0$. Moreover, let $F_1$ and $F_2$ denote the corresponding cumulative distribution functions (cdf's). We then define:
\begin{align*}
c_1 &= \int_{0}^{b} f_1(z) dz = F_1(b), \\[2mm]
c_2 &= \int_{b}^{\infty} f_2(z) dz = 1 - F_2(b).
\end{align*}
Then let $r \in [0, 1]$, and define a new pdf $f$ as follows,
\begin{equation}
\label{eq:thresholdMixture}
f(z) = I(z \leq b)\cdot r \cdot c_1^{-1} f_1(z) + I(z > b) \cdot (1 - r) \cdot c_2^{-1} f_2(z),
\end{equation}
where
\[
I(z \leq b) = \begin{cases}
1, & z \leq b\\
0, & z > b.
\end{cases}
\]
It is easy to  verify that $f$ is a legitimate pdf by checking that the integral of $f$ is $1$. 
\medskip
\noindent Assume in particular that $f_2(z) = 0$ for $z < b$. Thus, we have:
$$
c_2 = \int_{b}^{\infty} f_2(z) dz = \int_{0}^{\infty} f_2(z) dz  = 1.
$$
In this case \eqref{eq:thresholdMixture} can be written as:
\begin{equation}
\label{eq:thresholdMixtureC}
f(z) = I(z \leq b) \cdot r \cdot \frac{f_1(z)}{F_1(b)} + I(z > b) \cdot (1 - r) \cdot f_2(z) .
\end{equation}

\medskip

\noindent Moreover, we assume that $r = c_1=F_1(b)$. Then \eqref{eq:thresholdMixtureC} can be written as:
\begin{align}
\nonumber
\label{eq:thresholdMixtureD}
f(z) &= I(z \leq b) \cdot f_1(z) + I(z > b) \cdot (1 - c_1) \cdot f_2(z) \\[2mm]
&= I(z \leq b) \cdot f_1(z)  + I(z > b) \cdot (1-F_1(b)) f_2(z).
\end{align}
\noindent
Let $f_1= f_{\leq b}$ and $f_2=f_{> b}$ be the pdfs of the distributions below and above the threshold and $f_{\leq b|z\leq b}$ and $f_{> b|z > b}$ are the corresponding conditional pdfs of the distributions below and above the threshold, conditioning on $Z \leq b$ and $Z > b$, respectively. Then the pdf of the composite model according to \eqref{eq:thresholdMixtureD} is given by
\begin{align}
\begin{split}
f(z) = & I(z \leq b) f_{\leq b}(z) + (1-I(z \leq b)) (1-F_{\leq b}(b)) f_{> b}(z)\\
= & I(z \leq b) f_{\leq b|z\leq b}(z|z \leq b)F_{\leq b}(b)\\
& + (1-I(z \leq b)) f_{> b|z > b}(z|z > b)(1-F_{\leq b}(b)), \quad z > 0.
\end{split}
\label{eqn:comp_mod_1}
\end{align}
\cite{pickands1975} and \cite{balkema1974} have shown that $f_{> b|z > b}$, for the shifted variable $[Z-b|Z > b]$, may be approximated by a GPD for sufficiently large $b$. We restrict our attention to the Pareto type II distribution, which is a special case of the GPD, and in practice is very commonly used for extreme insurance claims. This means that 
\begin{equation}
f_{> b|z > b}(z|z > b) = \frac{\alpha/\beta}{\left(1+\frac{z-b}{\beta}\right)^{\alpha+1}}, \quad z > b,
\label{eqn:pareto}
\end{equation}
where $\alpha$ is a shape parameter and $\beta$ is a scale parameter. 

The bulk distribution is given by
\begin{equation}
f_{\leq b|z\leq b}(z|z \leq b) = \frac{f_{\leq b}(z)}{\Prob(Z \leq b)} = \frac{f_{\leq b}(z)}{F_{\leq b}(b)}, \quad 0 < z \leq b,
\label{eqn:bulk}
\end{equation}
where $f_{\leq b}$ may be any suitable distribution for small and moderate claims sizes, for instance the
Gamma, Weibull, log-normal or log-Gamma, that we have used in the simulation study.

As mentioned earlier, this model is usually fitted by first selecting the threshold $b$ with some methods. Then the data are divided according to $b$, fitting the bulk distribution \eqref{eqn:bulk} to the data below and the Pareto distribution \eqref{eqn:pareto} to the data above $b$.  We have chosen to do this with maximum likelihood estimation.

As an alternative, we consider the threshold mixing model of \cite{scollnik2007}. Their composite model consists of a truncated log-normal bulk distribution mixed with a Pareto type II distribution. The mixing is done in such a way that the resulting model \eqref{eq:thresholdMixture} is continuous in the threshold $b$, which is generally not the case for the model \eqref{eqn:comp_mod_1}. More specifically, the mixing weight
$r=\frac{\rho}{\rho+\beta}$ with $\rho=\sqrt{2\pi}\alpha b\sigma\Phi\left(\frac{\log(b)-\mu}{\sigma}\right)e^{\frac{1}{2}\left(\frac{
		\log(b)-\mu}{\sigma}\right)^{2}}$, where $\mu$ and $\sigma$ are parameters of log-normal distribution. The resulting model has a similar shape to log-normal distribution but with a thicker tail, which shows a better fit for larger losses in general insurance.  The authors have proposed a method for estimating the parameters $b$ and $\alpha$ with maximum likelihood, that we use in the simulation study.  This model here seems more advantageous than the model \eqref{eqn:comp_mod_1} since the parameters are estimated simultaneously given the data. 
	
	%However, our main purpose in this paper is to investigate the threshold selection methods and the influence of the bulk distribution given chosen threshold selection method, the model \eqref{eqn:comp_mod_1} loose the restriction on the choice of the threshold and in the sense, more flexible than the model of  \cite{scollnik2007}.

There are other threshold mixing models that might have been considered, among others the ones of \cite{Tancredi2006} or \cite{pigeon2011composite}. These are more flexible than \eqref{eqn:comp_mod_1}, but also more cumbersome to fit.

\section{Threshold selection methods}\label{sec:thresholdselect}

Choosing a plausible threshold for real life data is a classic tradeoff between bias and variance. A too large threshold yields few exceedances and consequently a larger uncertainty in the parameters of the GPD. On the other hand, if the threshold is too small, the GPD is no longer adequate for all the exceedances, which results in a larger bias in quantiles estimates. As earlier mentioned, a large number of threshold selection approaches has been proposed. As a large simulation study requires the threshold selection to be automatic, methods that involve visual inspection, such as the mean excess plot, are not included. Further, we have chosen to focus on methods that are easy to use and computationally not too demanding and also represent different types of approaches.

\subsection*{Heuristic methods}\label{subsec:heuristic}

The heuristic methods are essentially rules of thumb. They have the advantage of being very easy to implement. That is probably why they are so frequently used in practice by insurance companies. Let $z_{(1)} \leq \ldots \leq z_{(n)}$  be the data sorted in acsending order and $k$ be some real number in the interval $[0,n]$. Then, the methods included in the study all estimate the threshold $b$ by $\hat{b}=z_{([n-k])}$, where $[n-k]$ denotes the integer among $1,\ldots,n$ that is closest to $n-k$. \textbf{The fixed quantile rule} estimates the threshold by the $(1-\epsilon)\cdot 100\%$ empirical quantile of the data, for some fixed, level $\alpha$, which means that $k=\epsilon n$. \cite{DuMouchel} proposed to use $\epsilon=0.1$, but $\epsilon=0.05$ is more appropriate in our setting and is therefore used in the study. \textbf{The square root rule}, suggested by \cite{Ferreira2003} is to let $k=\sqrt{n}$. Finally, \textbf{the empirical rule} of \citep{LORETAN1994211} consists in letting $k=\frac{n^{2/3}}{\log(\log(n))}$.

\subsection*{Minimum AMSE of the Hill estimator}\label{subsec:damse}

This method is based on the Hill estimator of the tail index $\xi=\alpha^{-1}$, where $\alpha$ is the shape parameter of the Pareto distribution (see Section \ref{sec:compositemodels}). Using the fact that is $Z$ is Parteo distributed when $Z > b$, then $\ln(Z)-\ln(b)$ follows the exponential distribution with mean $\xi$,
\cite{hill1975} proposed the estimator
\[
\hat{\xi}_{k,n} \equiv \frac{1}{k}\sum_{i=1}^{k}\big(\ln z_{(n-i+1)}-\ln z_{(n-k)}\big),
\]
where $z_{(n-k)}$ is an estimate of the threshold $b$ and $k$ is now an integer . The idea of \cite{EVA2015} is to let $\hat{b}=z_{(n-\hat{k}_{0})}$ where $\hat{k}_{0}$ is an estimate of the value $k_{0}$ of $k$ that minimises the asymptotic mean squared error (AMSE) of $\hat{\xi}_{k,n}$, i.e. the sum of the squared bias and the variance of the asymptotic distribution of $\hat{\xi}_{k,n}$. The purpose is precisely to find a threshold that is a tradeoff between bias and variance. It may be shown that under certain regularity conditions (see \cite{EVA2015} for details)
\begin{align*}
\text{AMSE } \hat{\xi}_{k,n} &= \xi^2\left(\frac{1}{k} + \frac{\lambda^2}{(1-\rho)^{2}}\left(\frac{n}{k}\right)^{2\rho}\right),
\end{align*}
which is minimised for 
\[
k = k_{0} = \left\lfloor \left(\frac{(1-\rho)^2 n^{-2\rho}}{-2\rho \lambda^2}\right)^{1/(1-2\rho)}\right\rfloor \small,
\]
$\lfloor x \rfloor$  denoting the integer part of $x$. Here, $\rho$ and $\lambda$ are parameters that may be estimated as described in \cite{EVA2015}, resulting in 
\[
\hat{k}_{0} = \left\lfloor \left(\frac{(1-\hat{\rho})^2 n^{-2\hat{\rho}}}{-2\hat{\rho} \hat{\lambda}^2}\right)^{1/(1-2\hat{\rho})}\right\rfloor.
\]
This method is also available in the R-package \texttt{tea}.

\subsection*{Exponentiality test}\label{subsec:exptest}

As mentioned above, the log-differences $\ln(z_{(n-i+1)})-\ln(z_{(n-k)})$, $i=1,\ldots,k$ come (approximately) from an exponential distribution with mean $\xi$ if $k$ is large enough. Based on this, \cite{hill1975} proposed to choose $k$ as the minimum value for which the hypothesis of exponentiality does not fail. Since that tends to result in a too large $k$, \cite{Guillou2001} have suggested a modification that reduces this bias. More specifically, their estimate of $b$ is $\hat{b}=z_{(n-\hat{k})}$, where
\[
\hat{k} = \inf\{k:|Q_n(j)\geq 1.25, \quad\forall j \geq k\},
\]  
where
\begin{align*}
Q_n(k)= &\left\{\frac{1}{2\lfloor k/2 \rfloor+1}\sum_{j=k-\lfloor k/2 \rfloor}^{k+\lfloor k/2 \rfloor}T^2_n(j)\right\}^{1/2},\quad k\geq 1, k+\lfloor k/2 \rfloor <n\\
T_n(k) = &\sqrt{\frac{3}{k^3}}\frac{\sum_{i=1}^{k}(k-2i+1)U_i}{\frac{1}{k}\sum_{i=1}^{k}U_i}, \quad \text{ for } 1\leq k <n\\
U_i = &i\{\ln Z_{n-i+1:n}-\ln Z_{n-i:n}\}, \quad 1 \leq i \leq k \leq n.
\end{align*}
This method is also available in the \texttt{tea} package.
\subsection*{Gertensgarbe plot}\label{subsec:gertensgarbe}

The Gertensgarbe plot \citep{gertensgarbe1989} is inspired by the non-parametric Mann-Kendall test for monotonic trends in time series. Let $\Delta_{i}=z_{(i)}-z_{(i-1)}$, $i=2,\ldots,n$. The basic idea is that it is reasonable to expect that the behaviour of the $\Delta_{i}$s of a given dataset is different between the extreme and the normal observations. Hence, there should be a change point in the series of $\Delta_{i}$s, and this change point is considered as the starting point of the extreme region, and is therefore the estimate of the threshold $b$. To identify the change point, the test statistic of the sequential Mann-Kendall test is computed both for the $\Delta_{i}$s from $i=1$ to $i=n-1$ and for the differences in the reverse order. 
%More specifically, let applied twice, for the differences from start to the end of the dataset and vice versa. 
In this test, the normalized test series $U_i$ is given by
\[
U_i = \dfrac{\sum_{k=1}^{i}n_{k}-\frac{i(i-1)}{4}}{\sqrt{\frac{i(i+1)(2*i+5)}{72}}} \ \ \ \ \mbox{and} \ \ \ \ \tilde{U}_i = \dfrac{\sum_{k=1}^{i}\tilde{n}_{k}-\frac{i(i-1)}{4}}{\sqrt{\frac{i(i+1)(2*i+5)}{72}}},
\]
where $n_{k}=\sum_{j=1}^{k}I(\Delta_{j} < \Delta_{k})$ and $\tilde{n}_{k}=\sum_{j=1}^{k}I(\Delta_{n-j} < \Delta_{n-k})$. Then, the intersection point between the $U_i$s and the $\tilde{U}_i$s is the estimate $\hat{b}$. This method is for instance available in the R package $\texttt{tea}$.

\subsection*{Simultaneous estimation}\label{subsec:simest}

The last method we consider in the study is the simultaneous estimation of the threshold and other model parameters, assuming the composite model of \cite{scollnik2007} (see Section \ref{sec:compositemodels}). The main differences between this approach and the others, are that all the model parameters are estimated in one step instead of several, and that the bulk distribution is fixed as the log-normal.

\section{Simulation study}\label{sec:simstudy}

To investigate what the optimal threshold selection methods for non-life insurance data are and what the impact of the bulk distribution is, we have performed a simulation study where the sample size and tail properties of the bulk distribution are varied. Section \ref{subsec:paramset} gives the parameter settings for this study and the results are presented in Section \ref{subsec:results}.

\subsection{Parameter settings}\label{subsec:paramset}

Table \ref{table:param} and Table \ref{table:reserves} presents the choice of parameter values for each of the four
bulk distributions with corresponding standard deviations and 95\%, 99\% and 99.5\% reserves ($\epsilon = 0.05, 0.01$ and $0.005$, respectively). The corresponding non-truncated distributions have a mean of (approximately) $10$ and tail varying from moderate to quite heavy. Similarly, the parameters of Pareto distribution are set so that the mean claim size above the threshold is $50+b$ and the standard deviation is $118.3$, which distinguishes it from the body of the observations.  

As the claim sizes are assumed to be i.i.d., the number $\mathcal{N}_{\leq b}$ of claims below the threshold, given that the total number of claims is $\mathcal{N} = n_\mathrm{total}$, follows a binomial distribution with parameters $(n_\mathrm{total},p_{\leq b})$, where 
$p_{\leq b} = \Prob(Z \leq b) = F_{\leq b}(b;\B{\theta})$ is the probability that a claim is below the threshold, $\B{\theta}$ being the parameter vector of the bulk distribution. The probability $p_{\leq b}$ can be estimated either empirically or parametrically. The empirical estimator is simply the observed proportion of claims below the estimated threshold $\hat{b}$, i.e. 
$\hat{p}_{\mathrm{emp}}=n_{\leq \hat{b}}/n_\mathrm{total}$, with 
$n_{\leq \hat{b}} = \sum_{i=1}^{n_\mathrm{total}} I(z_{i} \leq \hat{b})$, and the parametric one is 
$\hat{p}_{\mathrm{the}} = F_{\leq b} (\hat{b}; \hat{\B{\theta}})$, $\hat{\B{\theta}}$ being the estimate of $\B{\theta}$. In this simulation study, both estimators will be applied. Further, claim sizes below the threshold are
simulated from the truncated bulk distribution $f_{\leq b|z \leq b}$ with the inversion method, i.e. by first sampling a uniform number $U^{*}$ and then transforming it with the inverse cdf $F_{\leq}^{-1}(bU^{*})$ of the truncated distribution, which is much more efficient than for instance rejection sampling.

The number of simulations in each experiment is $N = 1000$ and each experiment is performed as follows. One of the $4$ distributions is chosen to be the true bulk distribution, the true threshold $b$ is set as the $(1-\gamma)\cdot 100\%$ quantile of this distribution, where $\gamma$ is chosen as one of the values $0.08,0.06,0.04,0.02$. A sample $z_{1},\ldots,z_{n}$ of size $n$ is drawn from the composite distribution, where $n$ is varied between $n = 5000, 500, 50$, representing a large, medium, small sample size, respectively. 
For simplicity, $M_1$ to $M_6$ stand for the three heuristic methods, namely, the fixed quantile rule, the square root rule and the empirical rule, methods of minimum AMSE of the Hill estimator, exponentiality test and Gertensgarbe plot from Section \ref{sec:thresholdselect}, respectively. Then the above methods are applied to the sample. Subsequently, the parameters of each of the $4$ sub-threshold distributions and the Pareto above the threshold are estimated based on the sample and the threshold estimates. Thereafter, the method for simultaneous estimation of the threshold, denoted $M_7$, and model parameters, assuming a log-normal-Pareto composite distribution, is employed.  This results in the estimates $\hat{\bm{b}}_1,...,\hat{\bm{b}}_7$ and $\hat{\bm{\theta}}_1,...,\hat{\bm{\theta}}_{25}$.  As mentioned earlier, we only consider the classic Poisson distribution for the claim frequency of the portfolio, and the expected number of occurrences $\lambda$ is set to be $50$ or $500$. Finally, we estimate the reserve $q_{\epsilon}$, for $\epsilon=0.01$ and also $0.005$ and $0.05$, with $p_{\leq b}$ estimated both empirically and parametrically. This is done with Monte Carlo, as described in Algorithm \ref{alg:reserve}, using $m=1000000$ simulations. This results in the estimates $\hat{q}_{\epsilon,ij}$, $i=1,\ldots,N$, $j=1,\ldots,25$. The quality of these estimates is then evaluated by the finite sample bias $b_{j}=\frac{1}{N}\sum_{i=1}^{N}(\hat{q}_{\epsilon,ij}-q_\epsilon)$ and the root mean squared error $\mathrm{RMSE}_{j}=\sqrt{\frac{1}{N}\sum_{i=1}^{N}(\hat{q}_{\epsilon,ij}-q_\epsilon)^2}$.
%and the mean absolute error $\mathrm{MAE}_{j}=\frac{1}{N}\sum_{i=1}^{N}|\hat{q}^{*}_{\epsilon,ij}-q^*_\epsilon|$.

\begin{table}[H]
	\centering
	\begin{tabular}{clcc}
		\hline\hline
		& \textsc{Distr.} & \textsc{Par. val.} & \textsc{Std.} \\
		\multirow{3}{2cm}{\textsc{sub-threshold}}&
		$\mathrm{Gamma}(\alpha,\beta)$ & $(10,1)$ & 3.2\\
		&$\mathrm{L.-normal}(\xi,\sigma)$ & $(1.5,1.27)$ & 16.1\\
		&$\mathrm{Weibull}(\alpha,\beta)$ & $(2,11.28)$ & 5.2 \\
		&$\mathrm{L.-Gamma}(\alpha,\beta)$ & $(6, 3.04)$ & 24.6\\
		\hline
		\textsc{over-threshold}&
		$\mathrm{Pareto}(\alpha^p,\beta^p)$ & $(2.5,75)$ & 118.3 \\
		\hline\hline
	\end{tabular}
	\caption{Parameter values for the composite claim severity distributions used in the simulation study.}
	\label{table:param}
\end{table}	
\begin{table}[H]
	\centering
	\begin{tabular}{clcccccc}
		\hline\hline
		\textsc{Distr.}&  \textsc{thres.} &\multicolumn{6}{c}{\textsc{Reserves}} \\
		& & \multicolumn{3}{c}{$\lambda=50$} & \multicolumn{3}{c}{$ \lambda=500$}\\
		& &  $95\%$ & $99\%$  &  $99.5\%$& $95\%$ & $99\%$&$99.5\%$ \\
		\multirow{4}{*}{\textsc{Ga.-P.}} & $92\%$ &1093.48 &1520.20& 1780.10& 8256.43& 9315.76& 9927.54\\
		&$94\%$&993.37 &1368.81& 1598.66& 7616.00& 8564.33& 9146.73\\
		&$96\%$&882.56 &1199.03 &1398.06& 6941.02& 7741.35&8205.58\\
		&$98\%$ & 761.03 &999.21& 1151.88& 6215.40& 6819.07&7182.40\\
		\hline\hline
		\multirow{4}{*}{\textsc{L.N.-P.}} & $92\%$ &1069.48&1496.36& 1757.34& 7917.01& 8996.43& 9634.01\\
		&$94\%$& 981.21&1364.03& 1598.57& 7345.19& 8286.38& 8832.31\\
		&$96\%$& 883.53& 1207.18& 1407.96& 6760.14& 7568.38& 8038.75\\
		&$98\%$ & 774.14 &1013.31& 1169.04& 6142.23& 6751.80& 7095.33\\
		\hline\hline
		\multirow{4}{*}{\textsc{W.e.-P.}} & $92\%$ &1096.54 & 1523.86& 1781.97&8236.65& 9299.75& 9932.62\\
		&$94\%$&996.61 & 1377.13& 1607.39& 7599.40& 8542.67&9088.27\\
		&$96\%$& 889.35& 1206.87& 1406.49& 6941.59& 7751.75& 8214.07\\
		&$98\%$ &769.00 & 1007.11& 1154.60& 6228.34 &6830.34& 7186.15\\						
		\hline\hline
		\multirow{4}{*}{\textsc{L.G.-P.}} & $92\%$ & 1060.90& 1494.26& 1755.46& 7897.08& 8963.65& 9573.30\\
		&$94\%$& 968.61& 1348.72& 1586.34& 7300.87&  8244.35& 8815.37\\
		&$96\%$& 865.56& 1189.95& 1387.81& 6681.29& 7496.77&7964.82\\
		&$98\%$ & 751.92& 992.05& 1140.36& 6020.76& 6628.83& 6985.85\\		
		\hline\hline
	\end{tabular}
	\caption{$95\%$, $99 \%$ and  $99.5 \%$ reserves for composite distributions with different values of threshold and expected number of occurrences.}
	\label{table:reserves}
\end{table}

\begin{algorithm}[t]
	\caption{\label{alg:reserve}}
	Input: $\hat{\theta}$, $\hat\alpha^p,\hat\beta^p$, $\lambda$, $\hat{b}$, $\hat{p}_{\leq b}$, $\epsilon$
	\begin{algorithmic}[1]
		\For{$i = 1,\ldots,m$}
		\State $\mathcal{X}_{i}^{*} = 0$
		\State Draw $\mathcal{N^*} \sim Poisson(\lambda)$ 
		\State Draw $\mathcal{N}_{\leq b}^{*} \sim Binomial(\mathcal{N^*},\hat{p}_{\leq b})$
		\For{$j = 1,\ldots,\mathcal{N}_{\leq b}^{*}$}
		\State Draw $U^{*}\sim U(0,1)$
		\State $Z^{*} = F_{\leq b}(\hat{b}U^{*};\hat{\theta})$
		\State $\mathcal{X}_{i}^{*} = \mathcal{X}_{i}^{*}+Z^{*}$
		\EndFor
		\For{$j = 1,\ldots,\mathcal{N}-\mathcal{N}_{\leq b}^{*}$}
		\State Draw $Y^{*}$ from $Pareto(\hat{\alpha^p},\hat{\beta^p})$
		\State $Z^{*b}=Y^{*}+\hat{b}$
		\State $\mathcal{X}_{i}^{*} = \mathcal{X}_{i}^{*}+ Z^{*b}$
		\EndFor
		\EndFor
		\State Sort as $ \mathcal{X}_{(1)}^{*}\leq\ldots\leq\mathcal{X}_{(m)}^{*}$
		\State Return $\hat{q}_{\epsilon}= {X}_{((1-\epsilon)m)}^{*}$
	\end{algorithmic}
\end{algorithm}
\subsection{Results}\label{subsec:results}
By applying the 7 threshold selection methods, Table \ref{tab:bias_large_92_5000_99} to Table \ref{tab:bias_large_92_50_99}  show the bias of the reserve estimates when $\epsilon =0.01, \lambda=50$ and the probability $p_{\leq b}$ is estimated empirically, with varying sample size and threshold values. It shows that as the sample size changes, the performances of different methods vary a lot. When data are sufficient, either the exponentiality test ($M_5$) or the square root rule ($M_2$) performs best, except $M_5$ works slight better when the right tail of the data becomes more extreme, i.e., the true threshold value becomes larger.  However, the performances of the two other heuristic methods, the fixed quantile rule ($M_1$) and the empirical rule ($M_3$), are not as good as the square root rule ($M_2$), especially when the true threshold value is large. The problem with the heuristic methods, as explained in Section \ref{sec:thresholdselect}, is that the choice of $k$ only depends on the sample size. Therefore, with different threshold values, one would expect the performance of the heuristic methods vary a lot. However, that does not seem to be the case for the square root rule. It is somewhat surprising that the minimum AMSE of the Hill estimator ($M_4$) which constructed based on the asymptotic properties of the Hill estimator, has a poor performance when the sample size is large. The contradictory results may be due to the fact that $M_4$ underestimates the threshold value in most cases. The underestimated threshold might violate the assumption of Pareto family above the threshold and thus increase the model errors. Further, we see that the choice of the bulk distribution makes a difference on estimating the reserves. Generally, when the data are from a heavy-tailed distribution, for instance, the Log-normal or the Log-gamma, the choice of the bulk distribution given the estimated threshold matters a lot, in the sense that the differences in the bias estimates among bulk distributions vary greatly. The effect is further amplified when the threshold is estimated by methods with $M_4$, $M_5$ or the Gertensgarbe plot ($M_6$). This result may be explained by the fact that there is large difference between the distribution below and above the threshold in the study, which enlarges the impact of the bulk distribution. In addition, the influence also depends on the threshold selection methods. For instance, there are not great differences among the bulk distributions when the heuristic methods ($M_1, M_2$ and $M_3$)  are used. As the sample size decreases, the performance of each method becomes comparatively worse due to fewer exceedances above the estimated threshold, which leads to large estimation errors and consequently unrealistic reserve estimates, regardless of the types of the data and the tail properties. Instead, with limited data, one could resort to the simultaneous estimation ($M_7$) since it performs best among all the methods, though its performance is unsatisfactory when the data are sufficient. There are two likely causes for the relative poor performance of $M_7$ when the sample size is large. One is that $M_7$ assumes a continuous distribution at the threshold which is certainly not the case in the study. The other explanation for these results might be the assumption of a fixed family of bulk distribution that makes $M_7$ less flexible than the other methods.
Moreover, due to the increase of the estimation uncertainty, the influence of the bulk distribution seems to be smaller when the sample sizes are medium and small. The patterns in the RMSE estimates are similar to the ones for bias estimates, and are not shown here.

%in Appendix \ref{AppendixA}.
%The simultaneous estimation ($M_7$) performs not as good as others when the sample data are sufficient, but it indeed has a better performance when the true bulk distribution is log-normal which is as expected. Also method $M_7$ performs quite stable regardless of the variation of true threshold values.  Moreover, when $n=50$, In this case, one could use the simultaneous estimation instead since it shows acceptable results than any other methods. 
%the heuristic methods seem not that sensitive to the true bulk distribution, which means under these methods, the candidate bulk distributions give the similar reserve estimates once the true distribution is fixed. In the contrast, given methods $M_4, M_5$ and $M_6$, the differences of the bias estimates among each candidate bulk distribution vary a lot and particularly become even larger as data become more extreme. Moreover, the performance of minimum AMSE of the Hill estimator ($M_4$) and Gertensgarbe plot ($M_6$) fluctuate considerably according to different type of bulk data. In particularly, Gertensgarbe plot tends to do well when the true bulk distribution is not heavy-tailed and the sample size is large. for each threshold selection method such that the influence of bulk distribution might be ignored except method $M_4$, which surprisingly has a better performance than that when data are sufficient. 

Corresponding results for $\epsilon= 0.005$ are shown in Tables \ref{tab:bias_large_92_5000_995} to \ref{tab:bias_large_98_5000_995}. Since the pattern is similar to that for $\epsilon = 0.01$, the simulations for $n =500, 50$ are not shown. Now it indicates that $M_2$   performs best among all methods regardless of the true threshold level. Method $M_4$ has a poor performance, as for $\epsilon= 0.01$, but it performs better when the data are more extreme and the sample size is small. Due to the larger estimation error with smaller $\epsilon$, the bias of the reserve estimates becomes relatively larger than for $\epsilon = 0.01$. The influence of the bulk distribution however remains the similar pattern as for $\epsilon= 0.01$.  Further, it seems the variation between the estimates from different bulk distributions tends to be larger for data from moderated-tailed distribution. The results for $\epsilon=0.05$ are not shown here because they are similar  to the ones for $\epsilon=0.01$. As the expected claim number $\lambda$ increases, the total loss becomes a sum of a larger number of independent, identically distributed random variables. Its distribution should therefore become more and more similar to the normal distribution. In light of that, one would expect the location of the threshold, and in particular the choice of bulk distribution to matter less than for smaller $\lambda$. However, the results are similar to the ones for $\lambda = 50$ and the influence of bulk distribution does not seem smaller. The results are therefore not displayed here. A possible explanation for this might be that the total loss distribution is not quite normal yet because of the heavy tail of the claim size distribution, that makes the convergence slow. This finding also indicates that the claim size distribution has more influence on the reserve estimation than the claim frequency distribution.

%the total loss becomes the sum of a larger number of independent, identically distributed random variables. Its distribution should therefore become more and more similar to the normal distribution.
%In light of that, one would expect methods $M_4$ and $M_5$, based on their asymptotic properties to perform better for $\lambda =500 $ than for $\lambda = 50$, but it does not seem to be the case. 
As the probability $p_{\leq b}$ is estimated theoretically ($\hat{p}_{\mathrm{the}}$), the corresponding results when $\epsilon= 0.01$, $\lambda=50$ and $n=5000$ are shown in Tables \ref{tab:bias_large_92_5000_99_theo} to \ref{tab:bias_large_98_5000_99_theo}. Since the results are similar when the data become more extreme, only the threshold values equal to the 92 \% and the 98\% quantiles are shown here. The larger bias estimates indicate there is a big difference between $\hat{p}_{\mathrm{emp}}$ and $\hat{p}_{\mathrm{the}}$. It demonstrates that $\hat{p}_{\mathrm{emp}}$ gives a better performance than that with $\hat{p}_{\mathrm{the}}$ when data are sufficient. As $n$ becomes smaller, $\hat{p}_{\mathrm{emp}}$ tends to overestimate the exceedance above the threshold, which results in larger reserves. One would expect the performance with $\hat{p}_{\mathrm{the}}$ to be better, but it does not seem to be the case, and the performance is worse than that with $\hat{p}_{\mathrm{emp}}$. This may be explained by the larger estimation error under limited sample, then $\hat{p}_{\mathrm{the}}$ is close to zero and the corresponding reserve become much larger than the true one. Further, the choice of bulk distribution influences the bias estimates more than that with $\hat{p}_{\mathrm{emp}}$ even when data are sufficient. This makes sense because there are larger variations between $\hat{p}_{\mathrm{the}}$'s given different bulk distributions.
\begin{table}[H]\small
	\centering
	\begin{tabular}{ccccccccc}
		\hline\hline
		\textsc{True m.}&  \textsc{Distr.} &$\text{M}_1$ & $\text{M}_2$& $\text{M}_3$& $\text{M}_4$&$\text{M}_5$ &$\text{M}_6$&$\text{M}_7$\\
		\hline\hline
		\multirow{4}{*}{\textsc{Ga.-P.}} & \textsc{Ga.-P.}&19.46&15.41&22.23&168.53&-16.22&18.58&-\\
		&\textsc{L.N.-P.}&20.20&-5.11&13.89&257.17&-61.12&14.04&-\\
		&\textsc{We.-P.}&20.01&25.58&24.75&245.53&-0.41&20.20&-\\
		&\textsc{L.G.-P.} &20.11&-9.45&12.64&77.73&-66.41&13.13&-\\
		&\textsc{L.N.-P.}* &- &- &- & - & - & -&-650.57\\
		\hline
		\multirow{4}{*}{\textsc{L.N.-P.}} & 
		\textsc{Ga.-P.} &15.63&13.38&13.79&214.63&-16.31&166.69&-\\
		&\textsc{L.N.-P.}&17.89&-0.72&11.13&230.13&-46.08&-170.61&-\\
		&\textsc{We.-P.}&22.06&20.51&17.90&274.05&-10.12&192.47&-\\
		&\textsc{L.G.-P.} &26.18&15.42&21.80&194.59&-21.99&131.64&-\\
		&\textsc{L.N.-P.}* &- &- &- & - & - & -&131.73\\
		\hline
		\multirow{4}{*}{\textsc{We.-P.}} &
		\textsc{Ga.-P.} &17.48&10.87&17.37&579.89&5.38&14.53&-\\
		&\textsc{L.N.-P.}&31.81&13.82&28.89&597.50&-8.26&28.86&-\\
		&\textsc{We.-P.}& 20.99&20.26&20.98&595.21&16.31&15.50&-\\
		&\textsc{L.G.-P.} & 44.81&36.52&47.29&555.70&22.21&44.39&-\\	
		&\textsc{L.N.-P.}* &- &- &- & - & - & -&-250.01\\
		\hline
		\multirow{4}{*}{\textsc{L.G.-P.}} 
		& \textsc{Ga.-P.}&17.25 &8.21&13.63&1256.56&-12.89&222.34&-\\
		&\textsc{L.N.-P.}&17.29 &-15.02&5.29&1277.32&-56.59&223.59&-\\
		&\textsc{We.-P.}&21.07&18.14&20.34&1273.89&-1.77&445.90&-\\
		&\textsc{L.G.-P.} &20.19 &-14.84&7.18&1226.31&-56.95&226.41&-\\
		&\textsc{L.N.-P.}* &- &- &- & - & - & -&-220.36\\
		\hline\hline
	\end{tabular}
	\caption{Bias in the reserve estimates when $\epsilon=0.01$, $n=5000$, $\lambda=50$, the true threshold is chosen as the $92\%$ quantile and $p_{\leq b}$ is estimated empirically.}
	\label{tab:bias_large_92_5000_99}
\end{table}

\begin{table}[H]\small
	\centering
	\begin{tabular}{ccccccccc}
		\hline\hline
		\textsc{True m.}&  \textsc{Distr.} &$\text{M}_1$ & $\text{M}_2$& $\text{M}_3$& $\text{M}_4$&$\text{M}_5$ &$\text{M}_6$&$\text{M}_7$\\
		\hline\hline
		\multirow{4}{*}{\textsc{Ga.-P.}} & \textsc{Ga.-P.}&13.17&163.67&16.78&-142.88&5.49&18.39&-\\
		&\textsc{L.N.-P.}&14.37&142.29&14.72&-36.36&-22.33&15.12&-\\
		&\textsc{We.-P.}&57.60&173.88&15.17&-50.83&17.74&17.98&-\\
		&\textsc{L.G.-P.} &15.30&138.54&14.89&-249.42&-29.13&16.28&-\\
		&\textsc{L.N.-P.}* &- &- &- & - & - & -&-568.26\\
		\hline
		\multirow{4}{*}{\textsc{L.N.-P.}} & 
		\textsc{Ga.-P.} &5.48&151.24&4.78&68.27&-24.95&316.12&-\\
		&\textsc{L.N.-P.}&9.17&136.28&6.87&157.46&-45.17&319.92&-\\
		&\textsc{We.-P.}&10.06&157.69&9.04&167.64&-18.90&342.23&-\\
		&\textsc{L.G.-P.} &15.61&152.57&17.48&53.41&-19.55&282.13&-\\
		&\textsc{L.N.-P.}* &- &- &- & - & - & -&-67.92\\
		\hline
		\multirow{4}{*}{\textsc{We.-P.}} &
		\textsc{Ga.-P.} &13.88&157.43&18.46&218.15&-10.97&14.59&-\\
		&\textsc{L.N.-P.}&33.91&158.27&34.90&239.41&-16.92&30.49&-\\
		&\textsc{We.-P.}& 38.92&164.69&20.92&231.46&-2.74&14.96&-\\
		&\textsc{L.G.-P.} & -83.56&182.52&51.77&191.41&15.40&44.85&-\\	
		&\textsc{L.N.-P.}* &- &- &- & - & - & -&-290.38\\
		\hline
		\multirow{4}{*}{\textsc{L.G.-P.}} 
		& \textsc{Ga.-P.}&21.19&149.22&24.64&536.12&-1.17&177.21&-\\
		&\textsc{L.N.-P.}&21.50 &125.32&21.92&557.16&-33.02&177.92&-\\
		&\textsc{We.-P.}&32.31&159.71&29.10&552.79&9.45&202.26&-\\
		&\textsc{L.G.-P.} & 22.93&126.29&23.98&506.13&-32.05&177.48&-\\
		&\textsc{L.N.-P.}* &- &- &- & - & - & -&-283.85\\
		\hline\hline
	\end{tabular}
	\caption{Bias in the reserve estimates when $\epsilon=0.01$, $n=5000$, $\lambda=50$, the true threshold is chosen as the $94\%$ quantile and $p_{\leq b}$ is estimated empirically.}
	\label{tab:bias_large_94_5000_99}
\end{table}

\begin{table}[H]\small
	\centering
	\begin{tabular}{ccccccccc}
		\hline\hline
		\textsc{True m.}&  \textsc{Distr.} &$\text{M}_1$ & $\text{M}_2$& $\text{M}_3$& $\text{M}_4$&$\text{M}_5$ &$\text{M}_6$&$\text{M}_7$\\
		\hline\hline
		\multirow{4}{*}{\textsc{Ga.-P.}} & \textsc{Ga.-P.}&629.69&16.64&15.96&-310.14&-13.58&14.72&-\\
		&\textsc{L.N.-P.}&603.98&13.03&16.63&-196.29&-31.92&14.54&-\\
		&\textsc{We.-P.}&621.73&15.27&14.50&-212.04&-3.64&13.02&-\\
		&\textsc{L.G.-P.} &623.84&13.01&17.63&-427.07&-34.72&15.89&-\\
		&\textsc{L.N.-P.}* &- &- &- & - & - & -&-453.37\\
		\hline
		\multirow{4}{*}{\textsc{L.N.-P.}} & 
		\textsc{Ga.-P.} &517.31&14.81&18.47&138.96&-1.41&313.97&-\\
		&\textsc{L.N.-P.}&521.62&17.20&22.27&140.31&9.48&334.23&-\\
		&\textsc{We.-P.}&526.02&20.10&21.20&141.19&4.37&394.69&-\\
		&\textsc{L.G.-P.} &525.21&31.52&30.42&136.53&15.33&103.73&-\\
		&\textsc{L.N.-P.}* &- &- &- & - & - & -&-122.92\\
		\hline
		\multirow{4}{*}{\textsc{We.-P.}} &
		\textsc{Ga.-P.} &416.09&12.19&11.99&263.22&50.31&10.33&-\\
		&\textsc{L.N.-P.}&438.04&30.05&27.87&483.23&45.22&25.42&-\\
		&\textsc{We.-P.}& 423.39&13.97&11.05&372.02&53.65&11.42&-\\
		&\textsc{L.G.-P.} &362.81 &51.00&41.62&138.54&70.42&20.43&-\\	
		&\textsc{L.N.-P.}* &- &- &- & - & - & -&-274.48\\
		\hline
		\multirow{4}{*}{\textsc{L.G.-P.}} & 
		\textsc{Ga.-P.}& 915.27&4.52&9.06&545.16&0.05&485.57&-\\
		&\textsc{L.N.-P.}&914.20 &-0.06&9.11&691.30&-15.50&487.26&-\\
		&\textsc{We.-P.}&932.29&9.74&12.24&702.96&10.02&504.79&-\\
		&\textsc{L.G.-P.} & 919.90&2.88&11.59&307.42&-12.63&487.73&-\\
		&\textsc{L.N.-P.}* &- &- &- & - & - & -&-288.55\\
		\hline\hline
	\end{tabular}
	\caption{Bias in the reserve estimates when $\epsilon=0.01$, $n=5000$, $\lambda=50$, the true threshold is chosen as the $96\%$ quantile and $p_{\leq b}$ is estimated empirically.}
	\label{tab:bias_large_96_5000_99}
\end{table}

\begin{table}[H]\small
	\centering
	\begin{tabular}{ccccccccc}
		\hline\hline
		\textsc{True m.}&  \textsc{Distr.} &$\text{M}_1$ & $\text{M}_2$& $\text{M}_3$& $\text{M}_4$&$\text{M}_5$ &$\text{M}_6$&$\text{M}_7$\\
		\hline\hline
		\multirow{4}{*}{\textsc{Ga.-P.}} & \textsc{Ga.-P.}&869.65&13.33&557.71&-43.36&6.60&21.63&-\\
		&\textsc{L.N.-P.}&859.68&14.34&553.11&181.25&1.56&22.58&-\\
		&\textsc{We.-P.}&877.42&11.91&530.29&138.82&8.27&22.64&-\\
		&\textsc{L.G.-P.} &866.26&15.14&560.04&-340.45&0.79&23.49&-\\
		&\textsc{L.N.-P.}* &- &- &- & - & - & -&-285.42\\
		\hline
		\multirow{4}{*}{\textsc{L.N.-P.}} & 
		\textsc{Ga.-P.} &496.76&10.57&72.95&435.27&5.37&16.26&-\\
		&\textsc{L.N.-P.}&498.42&14.60&72.58&438.15&6.52&363.74&-\\
		&\textsc{We.-P.}&499.94&14.53&74.08&442.53&11.88&263.88&-\\
		&\textsc{L.G.-P.} &505.64&25.17&78.60&445.57&29.45&-320.46&-\\
		&\textsc{L.N.-P.}* &- &- &- & - & - & -&-132.59\\
		\hline
		\multirow{4}{*}{\textsc{We.-P.}} &
		\textsc{Ga.-P.} &676.98&12.28&338.03&279.87&7.37&469.99&-\\
		&\textsc{L.N.-P.}&695.70&28.57&342.18&301.74&24.57&492.43&-\\
		&\textsc{We.-P.}& 679.67&10.43&342.45&287.45&8.53&477.29&-\\
		&\textsc{L.G.-P.} & 629.05&42.35&387.93&239.75&55.57&158.39&-\\	
		&\textsc{L.N.-P.}* &- &- &- & - & - & -&-184.26\\
		\hline
		\multirow{4}{*}{\textsc{L.G.-P.}} &
		\textsc{Ga.-P.}&972.58&10.11&141.20&-104.26&4.27&324.86&-\\
		&\textsc{L.N.-P.}& 972.55&9.85&143.02&62.37&-2.95&326.67&-\\
		&\textsc{We.-P.}&973.16&14.16&145.89&31.98&12.86&440.40&-\\
		&\textsc{L.G.-P.} & 984.71&12.34&143.47&-312.71&2.62&331.22&-\\
		&\textsc{L.N.-P.}* &- &- &- & - & - & -&-203.77\\
		\hline\hline
	\end{tabular}
	\caption{Bias in the reserve estimates when $\epsilon=0.01$, $n=5000$, $\lambda=50$, the true threshold is chosen as the $98\%$ quantile and $p_{\leq b}$ is estimated empirically.}
	\label{tab:bias_large_98_5000_99}
\end{table}
\begin{table}[H]\small
	\centering
	\begin{tabular}{ccccccccc}
		\hline\hline
		\textsc{True m.}&  \textsc{Distr.} &$\text{M}_1$ & $\text{M}_2$& $\text{M}_3$& $\text{M}_4$&$\text{M}_5$ &$\text{M}_6$&$\text{M}_7$\\
		\hline\hline
		\multirow{4}{*}{\textsc{Ga.-P.}} & \textsc{Ga.-P.}&289.14 &290.45&216.71&190.34&280.97&232.32&-\\
		&\textsc{L.N.-P.}&289.20&293.74&218.24&264.35&291.79  &219.51&-\\
		&\textsc{We.-P.}&308.53&301.55&233.22&258.54&266.19 &232.69&-\\
		&\textsc{L.G.-P.} &290.97&286.49&225.66&120.51&271.59&223.59&-\\
		&\textsc{L.N.-P.}* &- &- &- & - & - & -&-642.01\\
		\hline
		\multirow{4}{*}{\textsc{L.N.-P.}} & 
		\textsc{Ga.-P.} &188.34 &260.71&361.63&292.71&421.66  &358.71&-\\
		&\textsc{L.N.-P.}&191.14 &263.29&373.51&309.59&481.55  &374.21&-\\
		&\textsc{We.-P.}&200.17 &266.94&432.03&303.37&444.46  &380.87&-\\
		&\textsc{L.G.-P.} &199.69&267.27&369.71&268.65&466.54&322.16&-\\
		&\textsc{L.N.-P.}* &- &- &- & - & - & -&-75.16\\
		\hline
		\multirow{4}{*}{\textsc{We.-P.}} &
		\textsc{Ga.-P.} &249.33 &367.93&365.35&620.49&430.79 &273.35&-\\
		&\textsc{L.N.-P.}&270.41 &382.90&452.79&627.59&530.80 &293.76&-\\
		&\textsc{We.-P.}& 265.12 &375.41&457.93&626.47&550.18 &285.66&-\\
		&\textsc{L.G.-P.} &244.38 &391.59&270.39&624.51&463.69&241.21&-\\	
		&\textsc{L.N.-P.}* &- &- &- & - & - & -&-312.18\\
		\hline
		\multirow{4}{*}{\textsc{L.G.-P.}} &
		\textsc{Ga.-P.}& 356.17&394.40&256.25&321.07&384.09  &259.35&-\\
		&\textsc{L.N.-P.}& 364.25&372.42 &257.59&337.84&382.45 &247.53&-\\
		&\textsc{We.-P.}&378.59&399.52&241.54&339.77&387.27  &258.66&-\\
		&\textsc{L.G.-P.} & 315.40&388.89&245.62&305.43&383.29&241.21&-\\
		&\textsc{L.N.-P.}* &- &- &- & - & - & -&-273.41\\
		\hline\hline
	\end{tabular}
	\caption{Bias in the reserve estimates when $\epsilon=0.01$, $n=500$, $\lambda =50$, the true threshold is chosen as the $92\%$ quantile and $p_{\leq b}$ is estimated empirically.}
	\label{tab:bias_large_92_500_99}
\end{table}

\begin{table}[H]\small
	\centering
	\begin{tabular}{ccccccccc}
		\hline\hline
		\textsc{True m.}&  \textsc{Distr.} &$\text{M}_1$ & $\text{M}_2$& $\text{M}_3$& $\text{M}_4$&$\text{M}_5$ &$\text{M}_6$&$\text{M}_7$\\
		\hline\hline
		\multirow{4}{*}{\textsc{Ga.-P.}} & \textsc{Ga.-P.}&763.36 &516.36 &911.70 &-260.26&814.19 &557.81  &-\\
		&\textsc{L.N.-P.}&756.69&612.59 &938.73&-68.68& 814.33  &598.54 &- \\
		&\textsc{We.-P.}&769.35&600.89&928.12 &-92.01 &837.67 &553.79  &-\\
		&\textsc{L.G.-P.} & 765.62&537.16& 912.02& 459.30&864.72&570.75&-\\
		&\textsc{L.N.-P.}* &- &- &- & - & - & -&-288.96\\
		\hline
		\multirow{4}{*}{\textsc{L.N.-P.}} & 
		\textsc{Ga.-P.} &314.76&214.25 &489.94&94.35&538.83     &  191.08&- \\
		&\textsc{L.N.-P.}& 316.69 &211.82& 400.17&375.97&526.23      &452.52 &-  \\
		&\textsc{We.-P.}& 320.49&212.10&415.03&288.36&580.81     &398.39&- \\
		&\textsc{L.G.-P.} & 328.34& 215.47&474.44&-235.42&541.34&-158.24&- \\
		&\textsc{L.N.-P.}* &- &- &- & - & - & -&-126.62\\
		\hline
		\multirow{4}{*}{\textsc{We.-P.}} &
		\textsc{Ga.-P.} &503.74&382.39&412.94&173.37&992.53   &922.16 &- \\
		&\textsc{L.N.-P.}&534.94& 383.45&516.70&392.91&927.15  &925.70 &- \\
		&\textsc{We.-P.}&533.68 & 392.15&479.24&318.00&934.58   &980.08 &-  \\
		&\textsc{L.G.-P.} &578.35 &378.30&466.15&-174.19&997.87&901.84&-\\	
		&\textsc{L.N.-P.}* &- &- &- & - & - & -&-183.30\\
		\hline
		\multirow{4}{*}{\textsc{L.G.-P.}} & 
		\textsc{Ga.-P.}&308.12 &497.40&432.46 &101.58 &962.43 &703.59&-\\
		&\textsc{L.N.-P.}& 327.16 & 524.81& 436.04 &72.45&1135.69 &732.40&-  \\
		&\textsc{We.-P.}& 311.01& 494.69&459.65&41.31&1630.23 &798.63&- \\
		&\textsc{L.G.-P.} & 328.38& 497.93& 441.76 &-359.11&1195.88&616.57&- \\
		&\textsc{L.N.-P.}* &- &- &- & - & - & -&-200.92\\
		\hline\hline
	\end{tabular}
	\caption{Bias in the reserve estimates when $\epsilon=0.01$, $n=500$, $\lambda=50$, the true threshold is chosen as the $98\%$ quantile and $p_{\leq b}$ is estimated empirically.}
	\label{tab:bias_large_98_500_99}
\end{table}
\noindent
\begin{table}[H]\small
	\centering
	\begin{tabular}{ccccccccc}
		\hline\hline
		\textsc{True m.}&  \textsc{Distr.} &$\text{M}_1$ & $\text{M}_2$& $\text{M}_3$& $\text{M}_4$&$\text{M}_5$ &$\text{M}_6$&$\text{M}_7$\\
		\hline\hline
		\multirow{4}{*}{\textsc{Ga.-P.}} & \textsc{Ga.-P.}&285.25&132.34&407.56&640.33&1020.67&2890.12 &-\\
		&\textsc{L.N.-P.}&573.24&887.34&677.45&959.25&42.25&4.27&-\\
		&\textsc{We.-P.}&462.09 &930.12&412.05&728.00&1160.12&4.08&-\\
		&\textsc{L.G.-P.} &589.25& 703.34&858.99& 1069.03& 474.69& 1159.34&-\\
		&\textsc{L.N.-P.}* &- &- &- & - & - & -&-0.06\\
		\hline
		\multirow{4}{*}{\textsc{L.N.-P.}} & 
		\textsc{Ga.-P.} &636.43&429.34&148.23&1020.59 & 931.23&16.03&-\\
		&\textsc{L.N.-P.}&772.67& 439.55&672.38&1450.39&502.38&16.19 &-\\
		&\textsc{We.-P.}&984.32& 432.32&595.29&17.45 &1383.45 &15.94&-\\
		&\textsc{L.G.-P.} &950.44&353.56&462.85&1932.22 &151.91& 18.31 &- \\
		&\textsc{L.N.-P.}* &- &- &- & - & - & -&-0.01\\
		\hline
		\multirow{4}{*}{\textsc{We.-P.}} &
		\textsc{Ga.-P.} &803.34 &780.01&818.34&997.23&1743.69&1070.89 &- \\
		&\textsc{L.N.-P.}&815.19& 661.23&552.24&602.59& 1237.84& 278.86&- \\
		&\textsc{We.-P.}& 724.25&620.11&781.19&160.08&1052.15 &1952.57&-\\
		&\textsc{L.G.-P.} &860.25& 701.33&532.31 &803.16&370.08 &305.09&-\\	
		&\textsc{L.N.-P.}* &- &- &- & - & - & -&-0.04\\
		\hline
		\multirow{4}{*}{\textsc{L.G.-P.}} &
		\textsc{Ga.-P.}&751.15&311.11&53.26&630.75 &910.22&124.07 &-\\
		&\textsc{L.N.-P.}&  673.39 &453.10&56.92&1443.13&204.80 &778.89 &-\\
		&\textsc{We.-P.}&848.24&507.98&58.88&2.37& 891.02&129.07 &-\\
		&\textsc{L.G.-P.} & 805.21&245.34&905.82&1360.11& 497.18 &903.56  &- \\
		&\textsc{L.N.-P.}* &- &- &- & - & - & -&-0.04\\
		\hline\hline
	\end{tabular}
	\caption{Bias($1\times 10^{-4}$) in the reserve estimates when $\epsilon=0.01$, $n=50$, $\lambda=50$, the true threshold is chosen as the $92\%$ quantile and the $p_{\leq b}$ is estimated empirically.}
	\label{tab:bias_large_92_50_99}
\end{table}
\begin{table}[H]\small
	\centering
	\begin{tabular}{ccccccccc}
		\hline\hline
		\textsc{True m.}&  \textsc{Distr.} &$\text{M}_1$ & $\text{M}_2$& $\text{M}_3$& $\text{M}_4$&$\text{M}_5$ &$\text{M}_6$&$\text{M}_7$\\
		\hline\hline
		\multirow{4}{*}{\textsc{Ga.-P.}} & \textsc{Ga.-P.} & 41.80&57.38 &52.06&619.47&139.89&45.01&-\\
		&\textsc{L.N.-P.}&43.13&35.35&45.84&710.89&99.03&40.65&-\\
		&\textsc{We.-P.}&45.41 &66.41&56.35&699.37&150.50&46.94&-\\
		&\textsc{L.G.-P.} & 42.01&32.01&43.04&535.17&92.40&39.13&-\\
		&\textsc{L.N.-P.}* &- &- &- & - & - & -&-858.04\\
		\hline
		\multirow{4}{*}{\textsc{L.N.-P.}} & 
		\textsc{Ga.-P.} &36.43&58.36&41.00&4257.23&113.86&3203.17&-\\
		&\textsc{L.N.-P.}& 40.04&43.67&36.67&4416.86&89.22&3242.84&-\\
		&\textsc{We.-P.}& 45.29&64.92&46.13&4373.06&131.38&3456.98&-\\
		&\textsc{L.G.-P.} & 47.60& 60.92&47.34&4047.25&109.63&2865.80&-\\
		&\textsc{L.N.-P.}* &- &- &- & - & - & -&383.20\\
		\hline
		\multirow{4}{*}{\textsc{We.-P.}} &
		\textsc{Ga.-P.} &40.29&50.27&46.58&13039.54&303.89&37.19&-\\
		&\textsc{L.N.-P.}& 54.78&50.75&58.63&13268.75&296.48&52.20&-\\
		&\textsc{We.-P.}& 45.27&58.78&50.49&13222.98&322.91&39.78&-\\
		&\textsc{L.G.-P.} & 68.09&75.37&76.56&12817.94&326.29&67.60&-\\	
		&\textsc{L.N.-P.}* &- &- &- & - & - & -&-258.52\\
		\hline
		\multirow{4}{*}{\textsc{L.G.-P.}} & \ \textsc{Ga.-P.} & 32.47& 35.40&32.98&29669.97&138.03&401.52&-\\
		&\textsc{L.N.-P.}& 33.13&12.81&24.92&29778.52&91.79&403.37&-\\
		&\textsc{We.-P.}& 35.38&45.57&39.94&29727.09&134.83&622.73&-\\
		&\textsc{L.G.-P.} & 34.82&12.18&28.26&29427.88&88.56&405.22&-\\
		&\textsc{L.N.-P.}* &- &- &- & - & - & -&-183.43\\
		\hline\hline
	\end{tabular}
	\caption{Bias in the reserve estimates when $\epsilon=0.005$, $n=5000$, $\lambda=50$, the true threshold is chosen as the $92\%$ quantile and $p$ is estimated empirically.}
	\label{tab:bias_large_92_5000_995}
\end{table}

\begin{table}[H]\small
	\centering
	\begin{tabular}{ccccccccc}
		\hline\hline
		\textsc{True m.}&  \textsc{Distr.} &$\text{M}_1$ & $\text{M}_2$& $\text{M}_3$& $\text{M}_4$&$\text{M}_5$ &$\text{M}_6$&$\text{M}_7$\\
		\hline\hline
		\multirow{4}{*}{\textsc{Ga.-P.}} & \textsc{Ga.-P.}&34.14&238.13 &45.93&25.67&203.46&50.75&-\\
		&\textsc{L.N.-P.}&36.04&218.54&45.64&130.10&170.78&49.47&-\\
		&\textsc{We.-P.}&79.90&247.55&45.93&116.84&212.82&51.03&-\\
		&\textsc{L.G.-P.} & 36.67&212.71&43.60&-81.10&173.48&47.20&- \\
		&\textsc{L.N.-P.}* &- &- &- & - & - & -&-762.49\\
		\hline
		\multirow{4}{*}{\textsc{L.N.-P.}} & 
		\textsc{Ga.-P.} &11.17&217.77&15.17&1277.48&67.27&6637.13&-\\
		&\textsc{L.N.-P.}& 14.73&201.09&16.75&1371.63&48.28&6682.18&-\\
		&\textsc{We.-P.}& 15.55&224.46&20.15&1388.56&71.18&6902.51&-\\
		&\textsc{L.G.-P.} & 22.06&218.07&27.50&1153.35&68.68&6311.60&-\\
		&\textsc{L.N.-P.}* &- &- &- & - & - & -&9.41\\
		\hline
		\multirow{4}{*}{\textsc{We.-P.}} &
		\textsc{Ga.-P.} &31.22&226.93&46.51&4703.31&86.04&36.83&-\\
		&\textsc{L.N.-P.}&50.73&226.65&61.51&4912.67&83.23 &51.96&-\\
		&\textsc{We.-P.}& 57.05&233.54&47.34&4838.82&95.94&37.57&-\\
		&\textsc{L.G.-P.} & -66.27&249.64&77.93&4431.04&111.19&66.15&-\\	
		&\textsc{L.N.-P.}* &- &- &- & - & - & -&-370.71\\
		\hline
		\multirow{4}{*}{\textsc{L.G.-P.}} & \ \textsc{Ga.-P.} & 36.28&205.63&49.24&12169.71&203.20&3602.27&-\\
		&\textsc{L.N.-P.}& 37.67&182.18&46.20&12402.55&177.44&3605.64&-\\
		&\textsc{We.-P.}& 45.57&215.12&53.25&12380.05&201.89&3839.03&-\\
		&\textsc{L.G.-P.} & 38.07&182.51&49.94&11863.79&166.67&3606.55&-\\
		&\textsc{L.N.-P.}* &- &- &- & - & - & -&-322.80\\
		\hline\hline
	\end{tabular}
	\caption{Bias in the reserve estimates when $\epsilon=0.005$, $n=5000$, $\lambda=50$, the true threshold is chosen as the $94\%$ quantile and $p_{\leq b}$ is estimated empirically.}
	\label{tab:bias_large_94_5000_995}
\end{table}

\begin{table}[H]\small
	\centering
	\begin{tabular}{ccccccccc}
		\hline\hline
		\textsc{True m.}&  \textsc{Distr.} &$\text{M}_1$ & $\text{M}_2$& $\text{M}_3$& $\text{M}_4$&$\text{M}_5$ &$\text{M}_6$&$\text{M}_7$\\
		\hline\hline
		\multirow{4}{*}{\textsc{Ga.-P.}} & \textsc{Ga.-P.}&6286.26&50.93&39.35&-81.83&74.51&38.97&-\\
		&\textsc{L.N.-P.}&6596.25&48.18&40.78&46.19&53.44&39.20&-\\
		&\textsc{We.-P.}&6043.28&50.18&35.38&31.80&83.56&36.68&-\\
		&\textsc{L.G.-P.} &6150.63&47.82&39.87&-226.00&48.37&38.86&-\\
		&\textsc{L.N.-P.}* &- &- &- & - & - & -&-625.75\\
		\hline
		\multirow{4}{*}{\textsc{L.N.-P.}} & 
		\textsc{Ga.-P.} &1038.01&39.61&35.87&1463.72&95.40&2331.64&-\\
		&\textsc{L.N.-P.}&1039.55 &40.17&39.37&1702.67&93.29&2532.43&-\\
		&\textsc{We.-P.}&1042.90 &44.04&38.12&1637.05&103.70&2582.86&-\\
		&\textsc{L.G.-P.} & 1050.14&55.47&46.99&1169.51&110.77&2000.20&-\\
		&\textsc{L.N.-P.}* &- &- &- & - & - & -&-127.31\\
		\hline
		\multirow{4}{*}{\textsc{We.-P.}} &
		\textsc{Ga.-P.} &5666.44&40.43&31.31&8238.99&1999.17&249.48&-\\
		&\textsc{L.N.-P.}&5867.97&57.35&47.02&8407.52&2879.72 &247.59&-\\
		&\textsc{We.-P.}& 5776.90&41.28&30.50&8445.95&3447.57&251.15&-\\
		&\textsc{L.G.-P.} & 5220.97&78.43&59.04&8144.35&2073.51&247.08&-\\	
		&\textsc{L.N.-P.}* &- &- &- & - & - & -&-391.17\\
		\hline
		\multirow{4}{*}{\textsc{L.G.-P.}} & \ \textsc{Ga.-P.} & 1928.26&22.10&23.04&600.96&465.51&3654.06&-\\
		&\textsc{L.N.-P.}& 1939.50&18.07&23.53&753.09&255.85&3664.82&-\\
		&\textsc{We.-P.}& 1951.90&28.41&25.61&728.27&386.65&3796.61&-\\
		&\textsc{L.G.-P.} & 1929.53&21.65&27.24&408.19&404.10&3663.42&-\\
		&\textsc{L.N.-P.}* &- &- &- & - & - & -&-376.79\\
		\hline\hline
	\end{tabular}
	\caption{Bias in the reserve estimates when $\epsilon=0.005$, $n=5000$, $\lambda=50$, the true threshold is chosen as the $96\%$ quantile and $p$ is estimated empirically.}
	\label{tab:bias_large_96_5000_995}
\end{table}

\begin{table}[H]\small
	\centering
	\begin{tabular}{ccccccccc}
		\hline\hline
		\textsc{True m.}&  \textsc{Distr.} &$\text{M}_1$ & $\text{M}_2$& $\text{M}_3$& $\text{M}_4$&$\text{M}_5$ &$\text{M}_6$&$\text{M}_7$\\
		\hline\hline
		\multirow{4}{*}{\textsc{Ga.-P.}} & \textsc{Ga.-P.}&7083.02&27.41&5355.30&165.84&95.83&39.67&-\\
		&\textsc{L.N.-P.}&7070.00&26.99&5212.65&388.57&83.46&41.52&-\\
		&\textsc{We.-P.}&6984.59&23.99&5828.47&344.53&98.29&41.55&-\\
		&\textsc{L.G.-P.} &6954.24&28.67&5961.84&-127.12&96.33&41.17&-\\
		&\textsc{L.N.-P.}* &- &- &- & - & - & -&-414.76\\
		\hline
		\multirow{4}{*}{\textsc{L.N.-P.}} & 
		\textsc{Ga.-P.} &1076.40&24.57&824.02&937.10&126.16&66.37&-\\
		&\textsc{L.N.-P.}&1080.18 &29.54&821.80&944.34&140.71&405.69&-\\
		&\textsc{We.-P.}&1080.06&28.44&822.47&945.80&138.21&308.69&-\\
		&\textsc{L.G.-P.} &1089.16&38.28 &926.98&948.19&137.99&-265.97&-\\
		&\textsc{L.N.-P.}* &- &- &- & - & - & -&-196.09\\
		\hline
		\multirow{4}{*}{\textsc{We.-P.}} &
		\textsc{Ga.-P.} &7768.76&29.28&6899.00&8719.35&143.68&1300.83&\\
		&\textsc{L.N.-P.}&7930.24&45.97&6864.99&8945.40&169.26&1378.91&-\\
		&\textsc{We.-P.}& 7762.26&28.62&6915.98&8857.29&144.40&1373.14&-\\
		&\textsc{L.G.-P.} & 7390.99&59.85&6176.68&8643.19&189.52&1009.05&-\\	
		&\textsc{L.N.-P.}* &- &- &- & - & - & -&-287.11\\
		\hline
		\multirow{4}{*}{\textsc{L.G.-P.}} & \ \textsc{Ga.-P.} &2753.60&29.49& 1762.35&-101.17&164.44&744.59&-\\
		&\textsc{L.N.-P.}&2747.32&27.26&1763.86&62.20&132.71&748.11& -\\
		&\textsc{We.-P.}&2749.46&32.02 &1766.72&33.82&168.26&859.25&-\\
		&\textsc{L.G.-P.} & 2746.19&31.55&1771.43&-306.56&158.46&751.18&-\\
		&\textsc{L.N.-P.}* &- &- &- & - & - & -&-291.88\\
		\hline\hline
	\end{tabular}
	\caption{Bias in the reserve estimates when $\epsilon=0.005$, $n=5000$, $\lambda=50$, the true threshold is chosen as the $98\%$ quantile and $p_{\leq b}$ is estimated empirically.}
	\label{tab:bias_large_98_5000_995}
\end{table}

\begin{table}[H]\small
	\centering
	\begin{tabular}{ccccccccc}
		\hline\hline
		\textsc{True m.}&  \textsc{Distr.} &$\text{M}_1$ & $\text{M}_2$& $\text{M}_3$& $\text{M}_4$&$\text{M}_5$ &$\text{M}_6$&$\text{M}_7$\\
		\hline\hline
		\multirow{4}{*}{\textsc{Ga.-P.}} & \textsc{Ga.-P.}&-451.95&-732.93 &-695.92&838.38&-674.14&-811.99&-\\
		&\textsc{L.N.-P.}&-447.92&-763.46&-706.62& 836.86&-734.95& -818.71  &-\\
		&\textsc{We.-P.}& -465.01&-706.06&-683.37&838.41&-640.13& -803.52 &-\\
		&\textsc{L.G.-P.} &-438.48&-767.47& -606.37&836.97&-743.33& -817.19&-\\
		&\textsc{L.N.-P.}* &- &- &- & - & - & -&-650.57\\
		\hline
		\multirow{4}{*}{\textsc{L.N.-P.}} & 
		\textsc{Ga.-P.} &-457.35&-739.19 &-710.06&956.01&-688.95&645.56 &-\\
		&\textsc{L.N.-P.}&-577.15&-700.46 &-697.71&981.89&-673.64  &886.18&- \\    
		&\textsc{We.-P.}&-410.46&-722.73& -699.75&996.50 &-674.16&1701.61 &-\\
		&\textsc{L.G.-P.} & -271.82& -475.08& -443.28&  989.36&-471.50&1703.31 &-\\
		&\textsc{L.N.-P.}* &- &- &- & - & - & -&131.73\\
		\hline
		\multirow{4}{*}{\textsc{We.-P.}} &
		\textsc{Ga.-P.} &-414.23&-726.32&-692.14&983.59&-670.13&-824.38&-   \\
		&\textsc{L.N.-P.}&-393.88&-681.55&-568.09&951.50&-649.26& -550.49 &- \\
		&\textsc{We.-P.}&  -439.80 &-709.31 &-686.14&971.50&-649.84& -823.27&-  \\
		&\textsc{L.G.-P.} & 88.67 & -454.87&-363.06&993.32&-451.43 &-162.46&-\\	
		&\textsc{L.N.-P.}* &- &- &- & - & - & -&-250.01\\
		\hline
		\multirow{4}{*}{\textsc{L.G.-P.}} & 
		\textsc{Ga.-P.}&-477.83&-747.99&-716.41&1214.78&-688.85&-231.31&-\\
		&\textsc{L.N.-P.}& -417.70 &-773.65&-715.77& 1119.66&-744.55& 48.70&- \\
		&\textsc{We.-P.}&-456.27&-724.30& -701.48&1121.16&-663.49& 815.42 &- \\
		&\textsc{L.G.-P.} & -310.36& -747.58&-672.77&1121.57&-722.14&  447.49&-\\
		&\textsc{L.N.-P.}* &- &- &- & - & - & -&-220.36\\
		\hline\hline
	\end{tabular}
	\caption{Bias in the reserve estimates when $\epsilon=0.01$, $n=5000$, $\lambda=50$, the true threshold is chosen as the $92\%$ quantile and $p_{\leq b}$ is estimated theoretically.}
	\label{tab:bias_large_92_5000_99_theo}
\end{table}

\begin{table}[H]\small
	\centering
	\begin{tabular}{ccccccccc}
		\hline\hline
		\textsc{True m.}&  \textsc{Distr.} &$\text{M}_1$ & $\text{M}_2$& $\text{M}_3$& $\text{M}_4$&$\text{M}_5$ &$\text{M}_6$&$\text{M}_7$\\
		\hline\hline
		\multirow{4}{*}{\textsc{Ga.-P.}} & \textsc{Ga.-P.}&700.97  &-322.11&745.54&668.43 &-294.24 &-152.39 & -\\
		&\textsc{L.N.-P.}&883.17 & -305.99 &794.98&631.53&-297.51&53.56 &-\\
		&\textsc{We.-P.}&837.88 &-326.66& 929.84&633.84&-286.40&-159.69&-\\
		&\textsc{L.G.-P.} &817.28&-283.94& 835.29 & 634.80&-292.21&119.54&-\\
		&\textsc{L.N.-P.}* &- &- &- & - & - & -&-285.42\\
		\hline
		\multirow{4}{*}{\textsc{L.N.-P.}} & 
		\textsc{Ga.-P.} &-200.38&-335.21&-262.96&-198.45&-295.92&-199.92  &-   \\
		&\textsc{L.N.-P.}&465.01& -133.97 &410.38& 519.48& -159.45&450.24 &-\\
		&\textsc{We.-P.}&-153.63&-337.10&-318.02& -104.23 &-294.36&490.77 &-   \\
		&\textsc{L.G.-P.} &533.17&143.75&792.34& 930.78& 68.16&491.490 &-\\
		&\textsc{L.N.-P.}* &- &- &- & - & - & -&-132.59\\
		\hline
		\multirow{4}{*}{\textsc{We.-P.}} &
		\textsc{Ga.-P.} &723.21&-241.83  &703.95 &648.18&-249.45&537.98&-  \\
		&\textsc{L.N.-P.}&716.16  &235.73&810.14&658.48& -54.11&813.13 &- \\
		&\textsc{We.-P.}&710.39& -317.69 & 874.91 &746.78&-277.25 &552.16 &-\\
		&\textsc{L.G.-P.} &775.84& 771.59&954.76&977.48&  256.30&924.05&-  \\	
		&\textsc{L.N.-P.}* &- &- &- & - & - & -&-184.26\\
		\hline
		\multirow{4}{*}{\textsc{L.G.-P.}} & 
		\textsc{Ga.-P.}&-164.77   &-342.93 &-271.05&548.84 &-308.64&-200.20&-   \\
		&\textsc{L.N.-P.}&552.42&-292.86&241.52 & 590.59 &-281.29&185.95&-  \\
		&\textsc{We.-P.}&553.22 & -339.98 &147.82 & 559.13&-299.19 &663.19 &-  \\
		&\textsc{L.G.-P.} &319.06&-198.26&  813.78&  644.08&-226.86&565.54&-\\
		&\textsc{L.N.-P.}* &- &- &- & - & - & -&-203.77\\
		\hline\hline
	\end{tabular}
	\caption{Bias in the reserve estimates when $\epsilon=0.01$, $n=5000$, $\lambda=50$, the true threshold is chosen as the $98\%$ quantile and $p_{\leq b}$ is estimated theoretically.}
	\label{tab:bias_large_98_5000_99_theo}
\end{table}

\section{Real data example}\label{sec:realdata}
In this section, we apply the composite models and the threshold selection methods considered in the simulation study to the Danish fire claim data. The data have been widely discussed in actuarial literature, \cite{scollnik2007} and \cite{actuarial}, among others. It consists of 2492 fire insurance losses in millions of Danish Kroner(DKK) from years 1980 to 1990 inclusive. %These data have been properly rescaled to reflect 1985 values. 
The data have a minimum value of 0.31 and a maximum value of 263.30. The mean and standard deviation are 3.06 and 7.98, respectively. The claim frequency distribution is set as Poisson distribution, with $\lambda=227$, which is the average number of claims per year in the historical data.  Table \ref{table:dk_para} gives us the parameter estimates for 4 composite models based on different threshold estimates. Moreover, the estimated reserve based on the parameter estimates are shown in Table \ref{table:reserve1}.

In general, the differences in estimated thresholds from the seven methods are quite large. We see that the minimum AMSE of the Hill estimator ($M_4$), the Gertensgarbe plot ($M_6$) and the simultaneous estimation ($M_7$) select a lower threshold than the others. Compared to the simulation results, one interesting finding is that the impact of the bulk distribution is minor for each threshold selection method at different solvency levels. As mentioned in Section \ref{sec:simstudy}, the large difference between the distributions below and above the threshold might be the reason why the bulk distribution influences the reserve estimates a lot. Therefore, the results in this case is likely to be related to less difference between the bulk distribution and the Pareto, though the ground truth is unknown. Further, it is shown that the results are similar to the ones from the simulation study, and might indicate that the reserves from the exponentiality test ($M_5$) and the square root rule ($M_2$) are the most trustworthy.

% In this example, the expected number of claims next year we assume might be ``large" so that the claim frequency and influence the could be approximated by a normal distribution. This may be one of the  reasons why bulk distribution makes no difference when estimating reserve within each threshold selection method. We try some smaller values, for instance, let $\lambda= 50, 20$. 

\begin{sidewaystable}[t]

	\centering
	\begin{tabular}{ccccccccccccc}
		\hline\hline
		\textsc{thres.est.}&\multicolumn{2}{c}{\textsc{over-thres. }}&\multicolumn{10}{c}{\textsc{bulk distr.}}\\
		\hline
		&\multicolumn{2}{c}{\textsc{Pa.($\hat{\alpha}^p, \hat{\beta}^p$)}} & \multicolumn{2}{c}{\textsc{GA.($\hat{\alpha}, \hat{\beta}$)}}& \multicolumn{2}{c}{\textsc{L.N.-P($\hat{\mu}, \hat{\sigma}$)}}& \multicolumn{2}{c}{\textsc{WE.($\hat{\alpha}, \hat{\beta}$)}}& \multicolumn{2}{c}{\textsc{L.G.($\hat{\alpha}, \hat{\beta}$))}}& \multicolumn{2}{c}{\textsc{L.N.($\hat{\alpha}, \hat{\beta}$)}*}\\
		$\hat{b}_1=8.41$ & 2.41 & 18.49 & 3.23 & 1.57 & 0.56 & 0.54 & 1.67& 2.33 & 9.22 & 8.81 & \multicolumn{2}{c}{-}\\
		$\hat{b}_2=17.07$ & 1.57 & 12.91 & 2.21 & 0.94 & 0.62 & 0.62 & 1.32 & 2.60 & 7.44 & 6.83 & \multicolumn{2}{c}{-}\\
		$\hat{b}_3=11.62$ & 1.94 & 14.39 & 1.76 & 1.27 & 0.58 & 0.57 & 1.51&2.44 & 8.43 & 7.92 &\multicolumn{2}{c}{-}\\
		$\hat{b}_4=2.46$  & 1.51 & 2.81 & 9.27 & 6.20 & 822.09 & 21.89 & 3.41 & 1.61 & 87.63 & 0.01 & \multicolumn{2}{c}{-}\\
		$\hat{b}_5=12.06$ &1.94 & 14.83 & 2.67 & 1.22 &0.59 & 0.58 & 1.48 & 2.46 &8.27 & 7.75 & \multicolumn{2}{c}{-}\\
		$\hat{b}_6=4.61$ &1.50 & 5.05 & 4.94 & 2.76 & 0.48 & 0.45 & 2.16 & 2.02 & 12.16 & 12.28 &  \multicolumn{2}{c}{-}\\
		$\hat{b}_7=1.44$ & 1.56& 0.36 &  \multicolumn{2}{c}{-}&  \multicolumn{2}{c}{-}&  \multicolumn{2}{c}{-}& \multicolumn{2}{c}{-} & 0.10& 0.18\\
		\hline\hline
		
	\end{tabular}
	\caption{Parameter estimates and threshold estimates for the composite models fitted to the Danish fire claim data.}
	\label{table:dk_para}
	\vspace{2\baselineskip}
			\small
	\begin{tabular}{lccccccccccccccc}
		\hline\hline
		\textsc{Thres.}&\multicolumn{15}{c}{\textsc{Esti.Reserves}}\\
		\hline
		&\multicolumn{3}{c}{\textsc{GA.-P}}&\multicolumn{3}{c}{\textsc{L.N.-P}}&\multicolumn{3}{c}{\textsc{WE.-P}}&\multicolumn{3}{c}{\textsc{L.G.-P}}&\multicolumn{3}{c}{\textsc{L.N.-P}*}
		\\
		& $95\%$ & $99\%$ & $99.5\%$& $95\%$ & $99\%$ & $99.5\%$& $95\%$ & $99\%$ & $99.5\%$& $95\%$ & $99\%$ & $99.5\%$& $95\%$ & $99\%$ & $99.5\%$\\
		$\hat{b}_1$ & 0.79 & 0.96 & 1.06&  0.78 & 0.95 & 1.05 &0.79 &0.96 &1.07 &0.78& 0.95& 1.05& \multicolumn{3}{c}{-}\\
		$\hat{b}_2$ &0.85&1.25 & 1.60 & 0.83 & 1.24 & 1.57 & 0.86 & 1.26 & 1.60 & 0.83 & 1.23 & 1.57&\multicolumn{3}{c}{-}\\
		$\hat{b}_3$ &0.82 & 1.08&1.27&0.81 & 1.06 & 1.25 & 0.83& 1.08 & 1.27 & 0.81 & 1.06 & 1.24&\multicolumn{3}{c}{-}\\
		$\hat{b}_4$&1.04&1.66&2.20&1.06 & 1.67 &2.21 & 1.04 & 1.64 & 2.18& 0.80 & 1.42 & 1.95&\multicolumn{3}{c}{-}\\
		$\hat{b}_5$ & 0.85 & 1.11 & 1.30 & 0.84 & 1.10 & 1.29 & 0.85 & 1.12 & 1.31 & 0.83 & 1.10 & 1.29&\multicolumn{3}{c}{-}\\
		$\hat{b}_6$& 1.01&1.66 & 2.19 & 1.00 & 1.65 & 2.21 & 1.01 & 1.65 & 2.19 & 1.00 & 1.64 & 2.22&\multicolumn{3}{c}{-}\\
		$\hat{b}_7$&\multicolumn{3}{c}{-}&\multicolumn{3}{c}{-}&\multicolumn{3}{c}{-}&\multicolumn{3}{c}{-}& 0.86 & 1.33 & 1.74\\
		\hline\hline
	\end{tabular}
	\caption{Reserve estimates (in billions of DKK) from different composite models and threshold selection methods for Danish fire claim data.}
	\label{table:reserve1}
	
\end{sidewaystable}

\section{Concluding remarks}\label{sec:conclusion}
The existence of large and extreme claims of a non-life insurance portfolio influences the ability of (re)insurers to estimate the reserve. The excess over-threshold method provides a way to capture and model the typical behaviour of insurance claim data. This paper discusses several composite models with the commonly used Gamma, Log-normal, Weibull and Log-gamma as bulk distributions, combined with a 2-parameter Pareto distribution above the threshold. We were interested in how the candidate threshold selection methods perform when  estimating the reserve and the influence of the bulk distribution, with varying sample size and tail properties. To investigate this, we performed a simulation study. 

Our study shows that when data are sufficient, the square root rule ($M_2$) has the overall best performance of reserve estimation among all approaches. The second best is the exponentiality test ($M_5$), especially when the right tail of the data is extreme. The two other heuristic methods, the fixed quantile rule ($M_1$) and the empirical rule ($M_3$), though lacking theoretical background, are comparatively better than the remaining ones. Moreover, the influence of the choice of bulk distribution on the reserve is large when the distribution is heavy-tailed. The effect is further amplified when the  threshold is estimated by methods with the minimum AMSE of the Hill estimator ($M_4$), the exponentiality test ($M_5$) or the Gertensgarbe plot ($M_6$). Thus, when these methods are used, the choice of bulk distribution should be carefully considered. As the sample size decreases, the simultaneous estimation ($M_7$) works best while its performance is unsatisfactory when the data are sufficient. One possible explanation for its poor performance under large sample is $M_7$ assumes a continuous distribution at the threshold which is certainly not the case in the study. The other  might be the assumption of a fixed family of bulk distribution that makes $M_7$ less flexible than the other methods. Due to the increase of the estimation uncertainty, the influence of the bulk distribution seems to be smaller when the sample sizes are medium and small. Furthermore, the effect of the bulk distribution seems larger as the probability $p_{\leq b}$ is estimated theoretically ($\hat{p}_{\mathrm{the}}$) rather than empirically ($\hat{p}_{\mathrm{emp}}$). This makes sense because there are larger variations between $\hat{p}_{\mathrm{the}}$'s given different bulk distributions. Despite the fact that $\hat{p}_{\mathrm{emp}}$ tends to overestimate the exceedance above the threshold when the sample size is small, the overall performance with $\hat{p}_{\mathrm{emp}}$ is better than that with $\hat{p}_{\mathrm{the}}$. This finding may indicate that  the empirical estimate of $p_{\leq b}$ is more robust than the theoretical one. 

In this study, we have restricted our attention to the choice of claim severity distributions, fixing the claim frequency distribution at the Poisson distribution. The reserve estimates can however easily be modified to account for other claim frequency distributions, such as the negative binomial. Further, it would be interesting to exploit more flexible simultaneous estimation by varying the bulk distribution and see whether the results improve, especially when the sample size is limited. 

%Last but not least, as the number of threshold selection methods grows, it would be interesting to see whether to further study should put more emphasis on ones that are either more simultaneous or much easier to be understood and applied by the (re)insurers.

\bibliography{ExtremesV2.1}

\begin{thebibliography}{25}
\expandafter\ifx\csname natexlab\endcsname\relax\def\natexlab#1{#1}\fi
\providecommand{\url}[1]{\texttt{#1}}
\providecommand{\href}[2]{#2}
\providecommand{\path}[1]{#1}
\providecommand{\DOIprefix}{doi:}
\providecommand{\ArXivprefix}{arXiv:}
\providecommand{\URLprefix}{URL: }
\providecommand{\Pubmedprefix}{pmid:}
\providecommand{\doi}[1]{\href{http://dx.doi.org/#1}{\path{#1}}}
\providecommand{\Pubmed}[1]{\href{pmid:#1}{\path{#1}}}
\providecommand{\bibinfo}[2]{#2}
\ifx\xfnm\undefined \def\xfnm[#1]{\unskip,\space#1}\fi
%Type = Article
\bibitem[{Balkema and de~Haan(1974)}]{balkema1974}
\bibinfo{author}{Balkema\xfnm[ A.]}, \bibinfo{author}{de~Haan\xfnm[ L.]}.
\newblock \bibinfo{title}{Residual life time at great age}.
\newblock \bibinfo{journal}{Annals of Probability}
  \bibinfo{year}{1974};\bibinfo{volume}{2}:\bibinfo{pages}{782--804}.
%Type = Book
\bibitem[{B{\o}lviken(2014)}]{bolviken_2014}
\bibinfo{author}{B{\o}lviken\xfnm[ E.]}.
\newblock \bibinfo{title}{Computation and Modelling in Insurance and Finance}.
\newblock International Series on Actuarial Science.
  \bibinfo{publisher}{Cambridge University Press}, \bibinfo{year}{2014}.
%Type = Book
\bibitem[{Caeiro and Gomes(2015)}]{EVA2015}
\bibinfo{author}{Caeiro\xfnm[ F.]}, \bibinfo{author}{Gomes\xfnm[ M.]}.
\newblock \bibinfo{title}{Threshold Selection in Extreme Value Analysis.}
\newblock Extreme value modeling and risk analysis: methods and applications.
  \bibinfo{publisher}{Wiley}, \bibinfo{year}{2015}.
%Type = Article
\bibitem[{Cebriaan et~al.(2003)Cebriaan, Denuit and Lambert}]{GPD}
\bibinfo{author}{Cebriaan\xfnm[ A.C.]}, \bibinfo{author}{Denuit\xfnm[ M.]},
  \bibinfo{author}{Lambert\xfnm[ P.]}.
\newblock \bibinfo{title}{Generalized pareto fit to the society of actuaries'
  large claims database.}
\newblock \bibinfo{journal}{North American Actuarial Journal}
  \bibinfo{year}{2003};\bibinfo{volume}{7}:\bibinfo{pages}{146--154}.
%Type = Book
\bibitem[{Coles(2001)}]{Coles2001}
\bibinfo{author}{Coles\xfnm[ S.]}.
\newblock \bibinfo{title}{An introduction to statistical modeling of extreme
  values}.
\newblock Springer Series in Statistics. \bibinfo{publisher}{Springer-Verlag},
  \bibinfo{year}{2001}.
%Type = Article
\bibitem[{Cooray and Ananda(2005)}]{actuarial}
\bibinfo{author}{Cooray\xfnm[ K.]}, \bibinfo{author}{Ananda\xfnm[ M.M.]}.
\newblock \bibinfo{title}{{Modeling actuarial data with a composite
  lognormal-Pareto model}}.
\newblock \bibinfo{journal}{Scandinavian Actuarial Journal}
  \bibinfo{year}{2005};\bibinfo{volume}{5}:\bibinfo{pages}{321--334}.
%Type = Article
\bibitem[{Davison and Smith(1990)}]{Davison.Smith1990}
\bibinfo{author}{Davison\xfnm[ A.C.]}, \bibinfo{author}{Smith\xfnm[ R.L.]}.
\newblock \bibinfo{title}{Models for exceedances over high thresholds}.
\newblock \bibinfo{journal}{Journal of the Royal Statistical Society}
  \bibinfo{year}{1990};\bibinfo{volume}{52}(\bibinfo{number}{3}):\bibinfo{pages}{393--442}.
%Type = Article
\bibitem[{Drees and Kaufmann(1998)}]{DREES1998149}
\bibinfo{author}{Drees\xfnm[ H.]}, \bibinfo{author}{Kaufmann\xfnm[ E.]}.
\newblock \bibinfo{title}{Selecting the optimal sample fraction in univariate
  extreme value estimation}.
\newblock \bibinfo{journal}{Stochastic Processes and their Applications}
  \bibinfo{year}{1998};\bibinfo{volume}{75}(\bibinfo{number}{2}):\bibinfo{pages}{149--172}.
%Type = Article
\bibitem[{DuMouchel(1983)}]{DuMouchel}
\bibinfo{author}{DuMouchel\xfnm[ W.H.]}.
\newblock \bibinfo{title}{Estimating the stable index {$\alpha$} in order to
  measure tail thickness: A critique}.
\newblock \bibinfo{journal}{The Annals of Statistics}
  \bibinfo{year}{1983};\bibinfo{volume}{11}(\bibinfo{number}{4}):\bibinfo{pages}{1019--1031}.
%Type = Article
\bibitem[{Ferreira et~al.(2003)Ferreira, de~Haan and Peng}]{Ferreira2003}
\bibinfo{author}{Ferreira\xfnm[ A.]}, \bibinfo{author}{de~Haan\xfnm[ L.]},
  \bibinfo{author}{Peng\xfnm[ L.]}.
\newblock \bibinfo{title}{On optimising the estimation of high quantiles of a
  probability distribution}.
\newblock \bibinfo{journal}{Statistics}
  \bibinfo{year}{2003};\bibinfo{volume}{37}(\bibinfo{number}{5}):\bibinfo{pages}{401--434}.
%Type = Article
\bibitem[{Gerstengarbe and Werner(1989)}]{gertensgarbe1989}
\bibinfo{author}{Gerstengarbe\xfnm[ F.]}, \bibinfo{author}{Werner\xfnm[ P.]}.
\newblock \bibinfo{title}{A method for the statistical definintion of
  extreme-value regions and their application to meteorological time series}.
\newblock \bibinfo{journal}{ZMeteorol}
  \bibinfo{year}{1989};\bibinfo{volume}{39}:\bibinfo{pages}{224--226}.
%Type = Article
\bibitem[{Gomes and Oliveira(2001)}]{Gomes2001}
\bibinfo{author}{Gomes\xfnm[ M.I.]}, \bibinfo{author}{Oliveira\xfnm[ O.]}.
\newblock \bibinfo{title}{{The bootstrap methodology in statistics of
  extremes--choice of the optimal sample fraction}}.
\newblock \bibinfo{journal}{Extremes}
  \bibinfo{year}{2001};\bibinfo{volume}{4}(\bibinfo{number}{4}):\bibinfo{pages}{331--358}.
%Type = Article
\bibitem[{Guillou and Hall(2001)}]{Guillou2001}
\bibinfo{author}{Guillou\xfnm[ A.]}, \bibinfo{author}{Hall\xfnm[ P.]}.
\newblock \bibinfo{title}{A diagnostic for selecting the threshold in extreme
  value analysis}.
\newblock \bibinfo{journal}{Journal of the Royal Statistical Society Series B}
  \bibinfo{year}{2001};\bibinfo{volume}{63}(\bibinfo{number}{2}):\bibinfo{pages}{293--305}.
%Type = Article
\bibitem[{Hall(1990)}]{HALL1990177}
\bibinfo{author}{Hall\xfnm[ P.]}.
\newblock \bibinfo{title}{Using the bootstrap to estimate mean squared error
  and select smoothing parameter in nonparametric problems}.
\newblock \bibinfo{journal}{Journal of Multivariate Analysis}
  \bibinfo{year}{1990};\bibinfo{volume}{32}(\bibinfo{number}{2}):\bibinfo{pages}{177--203}.
%Type = Article
\bibitem[{Hill(1975)}]{hill1975}
\bibinfo{author}{Hill\xfnm[ B.]}.
\newblock \bibinfo{title}{A simple general approach to inference about the tail
  of a distribu- tion}.
\newblock \bibinfo{journal}{Annals of Statistics}
  \bibinfo{year}{1975};\bibinfo{volume}{3}.
%Type = Book
\bibitem[{Kaas et~al.(2001)Kaas, Goovaerts, Dhaene and Denuit}]{kaas2001}
\bibinfo{author}{Kaas\xfnm[ R.]}, \bibinfo{author}{Goovaerts\xfnm[ M.]},
  \bibinfo{author}{Dhaene\xfnm[ J.]}, \bibinfo{author}{Denuit\xfnm[ M.]}.
\newblock \bibinfo{title}{Modern Actuarial Risk Theory -- Using R}.
\newblock \bibinfo{publisher}{Springer-Verlag Berlin Heidelberg},
  \bibinfo{year}{2001}.
%Type = Book
\bibitem[{Kleiber and Kotz(2003)}]{kleiber}
\bibinfo{author}{Kleiber\xfnm[ C.]}, \bibinfo{author}{Kotz\xfnm[ S.]}.
\newblock \bibinfo{title}{Statistical Size Distributions in Economics and
  Actuarial Sciences}.
\newblock \bibinfo{publisher}{Wiley}, \bibinfo{year}{2003}.
%Type = Book
\bibitem[{Klugman et~al.(2012)Klugman, Panjer and Willmot}]{klugman2012loss}
\bibinfo{author}{Klugman\xfnm[ S.]}, \bibinfo{author}{Panjer\xfnm[ H.]},
  \bibinfo{author}{Willmot\xfnm[ G.]}.
\newblock \bibinfo{title}{Loss Models: From Data to Decisions}.
\newblock Wiley Series in Probability and Statistics.
  \bibinfo{publisher}{Wiley}, \bibinfo{year}{2012}.
%Type = Article
\bibitem[{Loretan and Phillips(1994)}]{LORETAN1994211}
\bibinfo{author}{Loretan\xfnm[ M.]}, \bibinfo{author}{Phillips\xfnm[ P.C.]}.
\newblock \bibinfo{title}{Testing the covariance stationarity of heavy-tailed
  time series: An overview of the theory with applications to several financial
  datasets}.
\newblock \bibinfo{journal}{Journal of Empirical Finance}
  \bibinfo{year}{1994};\bibinfo{volume}{1}(\bibinfo{number}{2}):\bibinfo{pages}{211--248}.
%Type = Article
\bibitem[{Pickands(1975)}]{pickands1975}
\bibinfo{author}{Pickands\xfnm[ J.]}.
\newblock \bibinfo{title}{{Statistical Inference Using Extreme Order
  Statistics}}.
\newblock \bibinfo{journal}{The Annals of Statistics}
  \bibinfo{year}{1975};\bibinfo{volume}{3}(\bibinfo{number}{1}):\bibinfo{pages}{119--131}.
%Type = Article
\bibitem[{Pigeon and Denuit(2011)}]{pigeon2011composite}
\bibinfo{author}{Pigeon\xfnm[ M.]}, \bibinfo{author}{Denuit\xfnm[ M.]}.
\newblock \bibinfo{title}{{Composite Lognormal--Pareto model with random
  threshold}}.
\newblock \bibinfo{journal}{Scandinavian Actuarial Journal}
  \bibinfo{year}{2011};\bibinfo{volume}{2011}(\bibinfo{number}{3}):\bibinfo{pages}{177--192}.
%Type = Article
\bibitem[{Scarrott and Macdonald(2012)}]{Scarrott_areview}
\bibinfo{author}{Scarrott\xfnm[ C.]}, \bibinfo{author}{Macdonald\xfnm[ A.]}.
\newblock \bibinfo{title}{A review of extreme value threshold estimation and
  uncertainty quantification}.
\newblock \bibinfo{journal}{REVSTAT--Statistical Journal}
  \bibinfo{year}{2012};\bibinfo{volume}{10}(\bibinfo{number}{1}):\bibinfo{pages}{33--60}.
%Type = Article
\bibitem[{Scollnik(2007)}]{scollnik2007}
\bibinfo{author}{Scollnik\xfnm[ D.]}.
\newblock \bibinfo{title}{On composite lognormal-pareto models}.
\newblock \bibinfo{journal}{Scandinavian Actuarial Journal}
  \bibinfo{year}{2007};\bibinfo{volume}{1}:\bibinfo{pages}{20--33}.
%Type = Article
\bibitem[{Smith(1989)}]{smithRL}
\bibinfo{author}{Smith\xfnm[ R.L.]}.
\newblock \bibinfo{title}{Extreme value analysis of environmental time series:
  An application to trend detection in ground-level ozone}.
\newblock \bibinfo{journal}{Statistical Science}
  \bibinfo{year}{1989};\bibinfo{volume}{4}(\bibinfo{number}{4}):\bibinfo{pages}{367--377}.
%Type = Article
\bibitem[{Tancredi et~al.(2006)Tancredi, Anderson and O'Hagan}]{Tancredi2006}
\bibinfo{author}{Tancredi\xfnm[ A.]}, \bibinfo{author}{Anderson\xfnm[ C.]},
  \bibinfo{author}{O'Hagan\xfnm[ A.]}.
\newblock \bibinfo{title}{Accounting for threshold uncertainty in extreme value
  estimation}.
\newblock \bibinfo{journal}{Extremes}
  \bibinfo{year}{2006};\bibinfo{volume}{9}(\bibinfo{number}{2}):\bibinfo{pages}{87}.

\end{thebibliography}

\end{document}